\documentclass{article}
\usepackage{a4wide, cite, amsmath, graphics, subfigure}

\title{Glassy behaviour in a 3-state spin model}
\author{Lexie Davison\thanks{ldavison@thphys.ox.ac.uk} \hspace{4pt},
David Sherrington\thanks{d.sherrington1@physics.ox.ac.uk}, Juan P.
Garrahan\thanks{j.garrahan1@physics.ox.ac.uk} and Arnaud
Buhot\thanks{buhot@thphys.ox.ac.uk} \\ \textit{Theoretical Physics,
University of Oxford} \\ \textit{1 Keble Road, Oxford OX1 3NP,
UK}}

\date{\today}
\begin{document}
\maketitle

\begin{abstract}
In this article we study a simple spin model which has a
non-interacting Hamiltonian but constrained dynamics. The model, which
is a simplification of a purely toplogical cellular model
\cite{paper1,asteorig}, displays glassy behaviour, involves
activated processes and exhibits two-step relaxation. This is a
consequence of the existence of annihilation-diffusion processes on
two distinct time-scales, one temperature independent and the other an
exponential function of inverse temperature. In fact, there are several
such inter-coupled microscopic processes and great richness
therein. Two versions of the model are considered, one with a single
absorbing ground state and the other with a highly degenerate ground
state. These display qualitatively similar but quantitatively
distinct macroscopic behaviour and related, but different, microscopic
behaviour.
\end{abstract}

\section{Introduction}
In a recent paper Davison and Sherrington \cite{paper1} focussed on a purely
topological tiling model \cite{asteorig} which exhibited glassy
dynamical behaviour. This was driven by a desire to
investigate the behaviour of supercooled liquids in a model which
contained as few parameters as possible, but still displayed the
relevant physics. In this paper we continue the minimalistic
approach, and consider systems which are conceptually similar to the
topological model, but have the major advantage of involving variables
(spins) based on a fixed lattice. 

The original model was that of a two-dimensional topological froth,
constructed by tiling the plane using three-fold vertices only. However, unlike soap froths, the energy was determined by the
deviation of the cell topologies from a perfect hexagonal tiling:
$E=\sum_i ^N (n_i -6)^2$, where $n_i$ is the number of sides of cell
$i$, and the system evolved solely through stochastic T1-micro-dynamics
with Glauber-Kawasaki probabilities. We found that there was two-step
relaxation at low temperatures, and formed the following conceptual picture
describing the evolution of the system. At low temperatures the system
consists mainly of six-sided cells, and there are two
processes dominating the behaviour: on a fast
time-scale, pairs of pentagon-heptagon defects diffuse freely through the
hexagonal background, and on a slower time-scale isolated defects
absorb or create pairs of defects. The latter can be considered to be
an activated process, as it is energetically unfavourable for such a
pair of defects to be created. This conceptual framework of two
different processes leading to both fast and slow dynamics is
directly applicable to the lattice-based spin model with which this
paper is concerned, and thus one might expect it to yield qualitatively
similar results to the topological model. Given that this spin model is
computationally simpler and more tractable, one might
hope to be able to probe more deeply, and to investigate to what
extent one finds the same features displayed by other kinetically
constrained models (for examples see \cite{fredrikson,
eisinger,kurchan,  sollich, RitortQ, juanpe,barratloreto, padilla, theo}). In fact, as we show below, these expectations are borne
out and it is possible to provide a physical understanding of the
features observed. It is also both possible and instructive to
consider a generalization with a very different (highly degenerate)
ground state, which we cover in the latter half of the paper.

\section{The Model}
\begin{figure}[t!]
\begin{center}
\subfigure[The T1 moves for the topological
froth.]{\label{t1move}\resizebox{!}{140pt}{\includegraphics{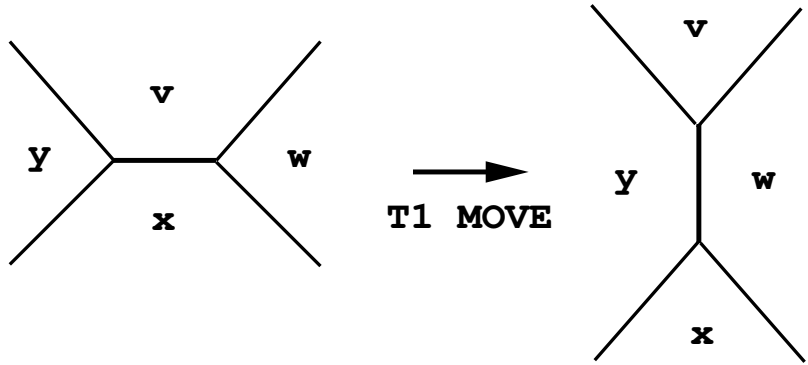}}}
\subfigure[The spin-flip rules for the present model: the total spin
is conserved and the choice of whether to attempt the upper or lower
signs is made randomly at each time-step. The quadruplet of cells is
identified uniquely by the dashed edge $e$.]{\label{spinmove}\resizebox{!}{176pt}{\includegraphics{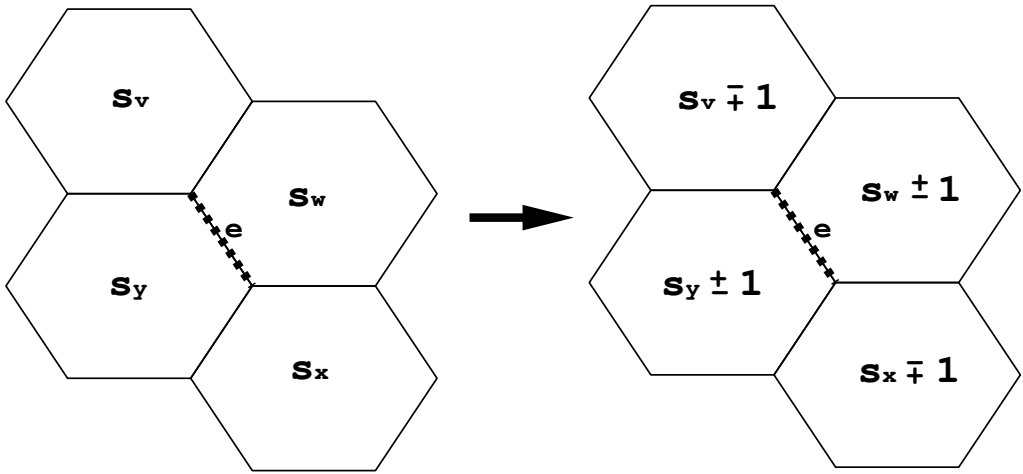}}}
\caption{\textbf{Spin-flip rules for the 3-state spin model and the topological
froth.} \label{moves}}
\end{center}
\end{figure}
The model we use comprises a perfect hexagonal tiling of the plane,
with a variable (spin) associated with each hexagon. 
The spin in cell $i$, denoted by $s_i$,
is restricted to the values $0,\pm 1$. This is analogous to
a topological froth in which the cells are all pentagons, hexagons or
heptagons, so that the topological charge $q_i=(6-n_i)$ of each cell is
restricted to the values $0, \pm 1$. However, the topological froth
model has a non-periodic and dynamically changing cell structure,
whereas this model is firmly fixed on a perfect hexagonal tiling.

We define the energy as follows:
\begin{equation}
\label{energy}
E= D \sum_{i=1}^N s_i ^2
\end{equation}
where $N$ is the total number of cells/spins in the system, and $D$
may be positive or negative. The case of $D>0$ emulates the original
topological model where hexagonal order is energetically preferable,
but $D<0$ is also of interest, as we shall show. The dynamics conserve
the total spin of the system and we choose our starting configuration
such that the total spin is always zero i.e.  
\begin{equation}
\label{cons}
\sum_{i=1}^N s_i = 0
\end{equation}
in analogy with the Euler rule $\sum_i ^N (6-n_i)=$0 which applies to
a froth \cite{euler1,euler2}. Therefore the ground state for $D>0$ consists of $s_i=0$ for all spins,
whereas for $D<0$ the ground state is degenerate, with half the spins
taking the value $+1$ and half taking the value $-1$.

The system is allowed to evolve through local spin-flips which are
similar to the T1 moves performed on topological froths. In the latter
two adjacent
cells have their topological charges decreased by $1$, and their two
common neighbours have their topological charges increased by
$1$: see Figure \ref{t1move}. In the present model the allowed
move-sets consist of choosing a pair of neighbouring cells and either
increasing their spins by 1 unit, and decreasing those of their common
neighbours by 1 unit, or vice versa; both possibilities need to be
allowed to avoid chirality inhibiting movement of spins throughout
the system. 

The actual dynamical process is as follows. At each time step an edge
is chosen randomly on the hexagonal lattice; this defines a set of
four cells as shown in Figure \ref{spinmove}. A choice is then made
randomly of whether to consider 
increasing the spins of the adjacent cells $y$ and $w$ (and thus decreasing $v,
x$) or vice versa, with equal probabilities for both cases. The
probability of actually performing the
move is dependent on the energy change that would be incurred, and is
given by a temperature-dependent
Metropolis-Kawasaki \footnote{We choose to use the Metropolis algorithm rather than Glauber
dynamics as in previous work because the qualitative features of the
results show no dependence on which of these algorithms we choose, and
Metropolis is the faster of these two. The reference to Kawasaki is
included to emphasise the fact that although more than one spin is
flipped at once, the total spin is conserved.} algorithm. Specifically, assuming $w$ and $y$
have been chosen as candidates for an increase in spin, the energy change associated with these spin-flips on spins $s_v,
s_w, s_x, s_y$ is:
\begin{equation}
\label{delta}
\Delta E(s_w,s_y;s_v,s_x)=2D(2+s_w + s_y - s_v - s_x)
\end{equation}
and the probability P of actually performing this move is:
\begin{equation}
\label{prob}
P(s_w,s_y;s_v,s_x)=(1-\delta_{s_w,1})(1-\delta_{s_y,1})(1-\delta_{s_v,-1})(1-
\delta_{s_x,-1}) \hspace{3pt} Min[1,\exp(-\beta
\Delta E(s_w,s_y;s_v,s_x)) ]
\end{equation}
where $\beta$ is the inverse temperature. The $\delta$-functions
ensure that the spins are forbidden to take values other than $\pm
1$ or 0. 

The simple form of equation (\ref{energy}), with no interaction between
the cells, shows that this system is thermodynamically trivial in
equilibrium and all the static equilibrium properties are readily
calculable. However, the microscopic dynamics are constrained and
non-trivial, involving several spins simultaneously; this leads to
glassy macro-dynamics. 

Most of the data results from simulations on a system of size N=9900,
although, in order to perform more accurate fits, in certain cases the
system size was increased to 160000. However, unless otherwise stated,
one should assume the former system size is in use. Periodic boundary
conditions are enforced in all cases.

This paper is structured in the following manner: first we present results for $D>0$, including a discussion on
the processes involved in relaxation of the system; this is followed with
results for $D<0$. Finally there is a more general discussion.

\subsection*{Brief review of the topological froth model}
As one of the aims of this paper is to show that the behaviour of
this model is indeed qualitatively very similar to that of the topological
froth, we shall briefly review the results for the topological froth
before proceeding \cite{asteorig, paper1}. 

In simulations in which the system is cooled at a variety of different
rates, one finds strong dependence of the energy on the cooling rate,
with the system unable to attain equilibrium within any
reasonable time-scale at very low temperatures. Measurements of a
two-time auto-correlation function show evidence of two-step relaxation,
with plateaux developing as the temperature is reduced. There are also
clear signs of aging when the system is not in equilibrium. When the
equilibrium  correlation functions are rescaled by a suitably defined
relaxation time, one finds that they collapse onto a master curve in the
late-$\beta$ relaxation regime, and that this master curve can be
fitted by a von Schweidler law as predicted by Mode-coupling
Theory (MCT) \cite{gotzerev}. The relaxation time is well-described by an offset
Arrhenius law, indicating strong glassy behaviour \cite{angell}. Although MCT in
fact predicts a power law, this does not fit the data particularly
well, and neither does the oft-used Vogel-Fulcher law.

By measuring a suitable temporal response function and plotting it
parametrically against the appropriate correlation function (starting from a
non-equilibrium configuration), the Fluctuation-Dissipation Theorem (FDT) is found to be upheld for times somewhat
longer than required for the onset of the correlation function
plateaux. For longer times FDT ceases to hold. After it is broken, one
observes non-monotonic behaviour, a feature which has also been noted
in several other models which can be considered to involve activation
over energy barriers \cite{fredrikson,
eisinger,kurchan,  sollich, RitortQ, juanpe,barratloreto, granular}.

\begin{figure}
\begin{center}
\subfigure[The behaviour of the energy with
temperature under slow cooling $(D>0)$. The values of $t_w$ are the waiting
times at each point.]{\label{hcool}\resizebox{!}{260pt}{\includegraphics{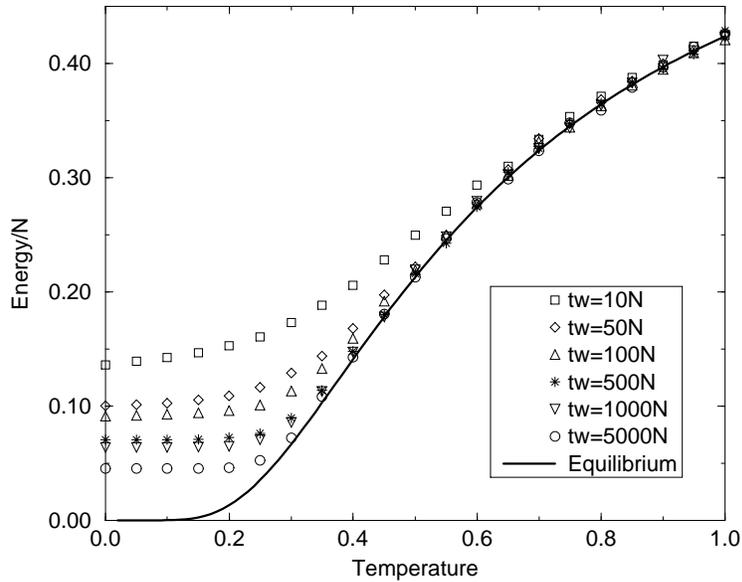}}}
\subfigure[The behaviour of the energy with
temperature after a rapid quench
$(D>0)$. The values of $t_w$ are the times, subsequent to the quench,
at which the energy is measured.]{\label{hquench} \resizebox{!}{260pt}{\includegraphics{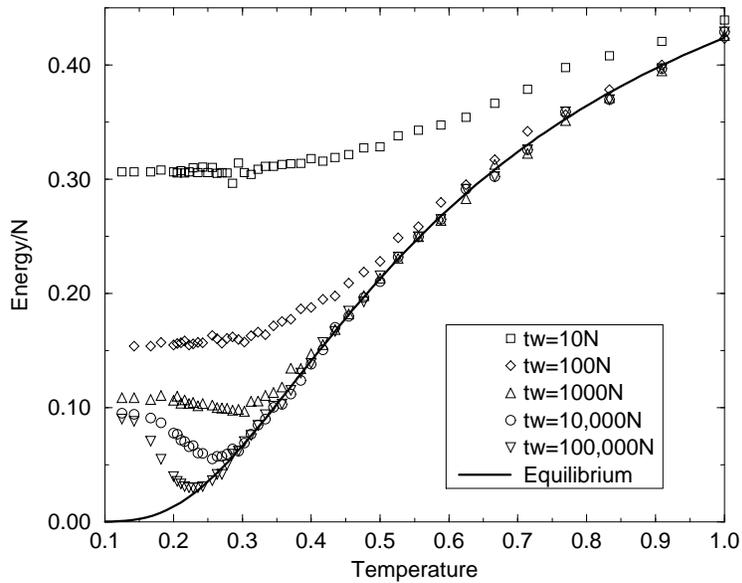}}}
\caption{\textbf{Energy against temperature for slow cooling and rapid
quench.}}
\end{center}
\end{figure}
\begin{figure}
\begin{center}
\subfigure[Energy against time. The curves for $\beta \leq 5$ reach
their equilibrium values, whereas the others do not.]{\label{enrgy}\resizebox{!}{270pt}{\includegraphics{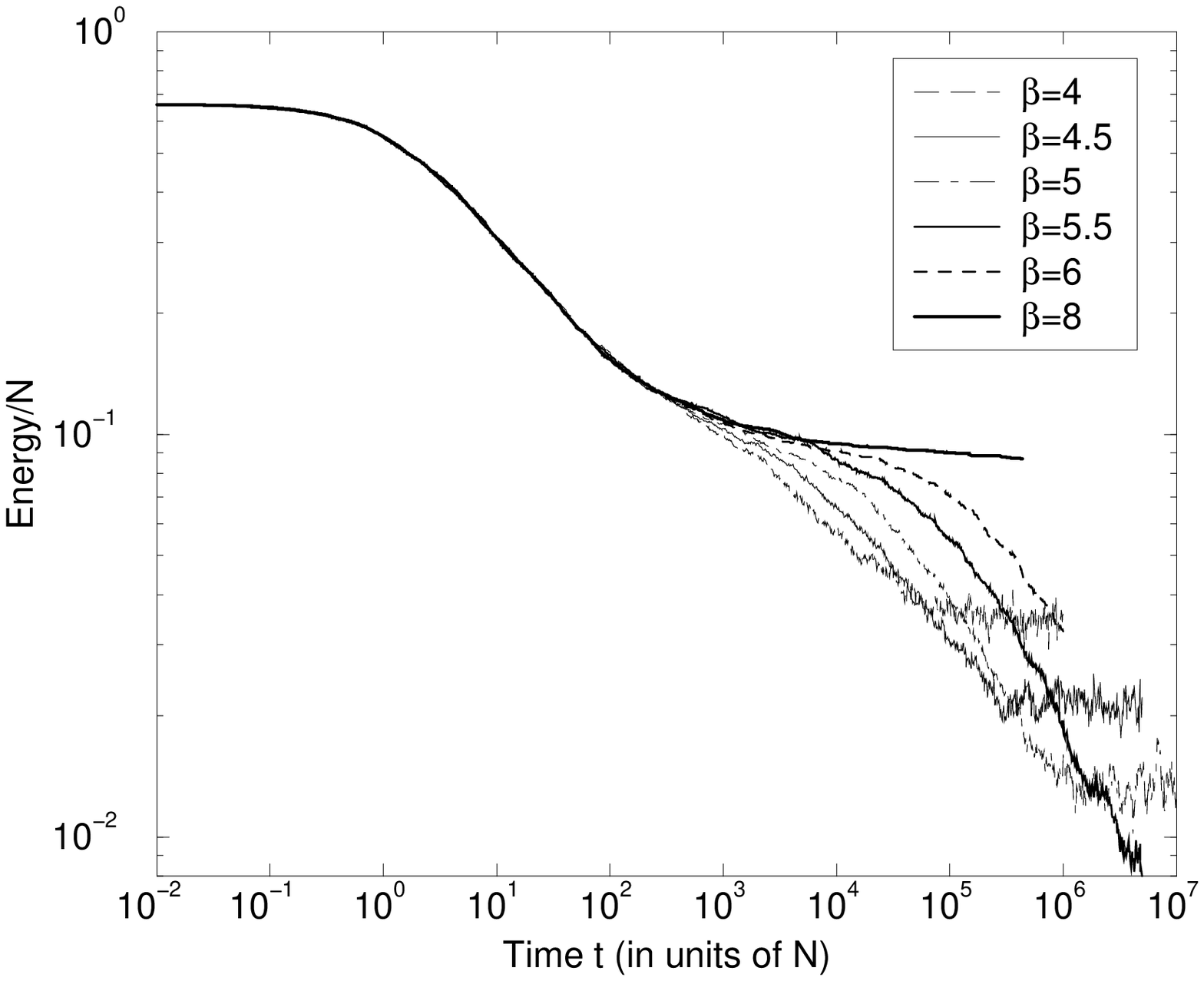}}}
\subfigure[Energy against $T \ln t$ (where $t$ is measured in units of $N$).]{\label{lognrj}\resizebox{!}{270pt}{\includegraphics{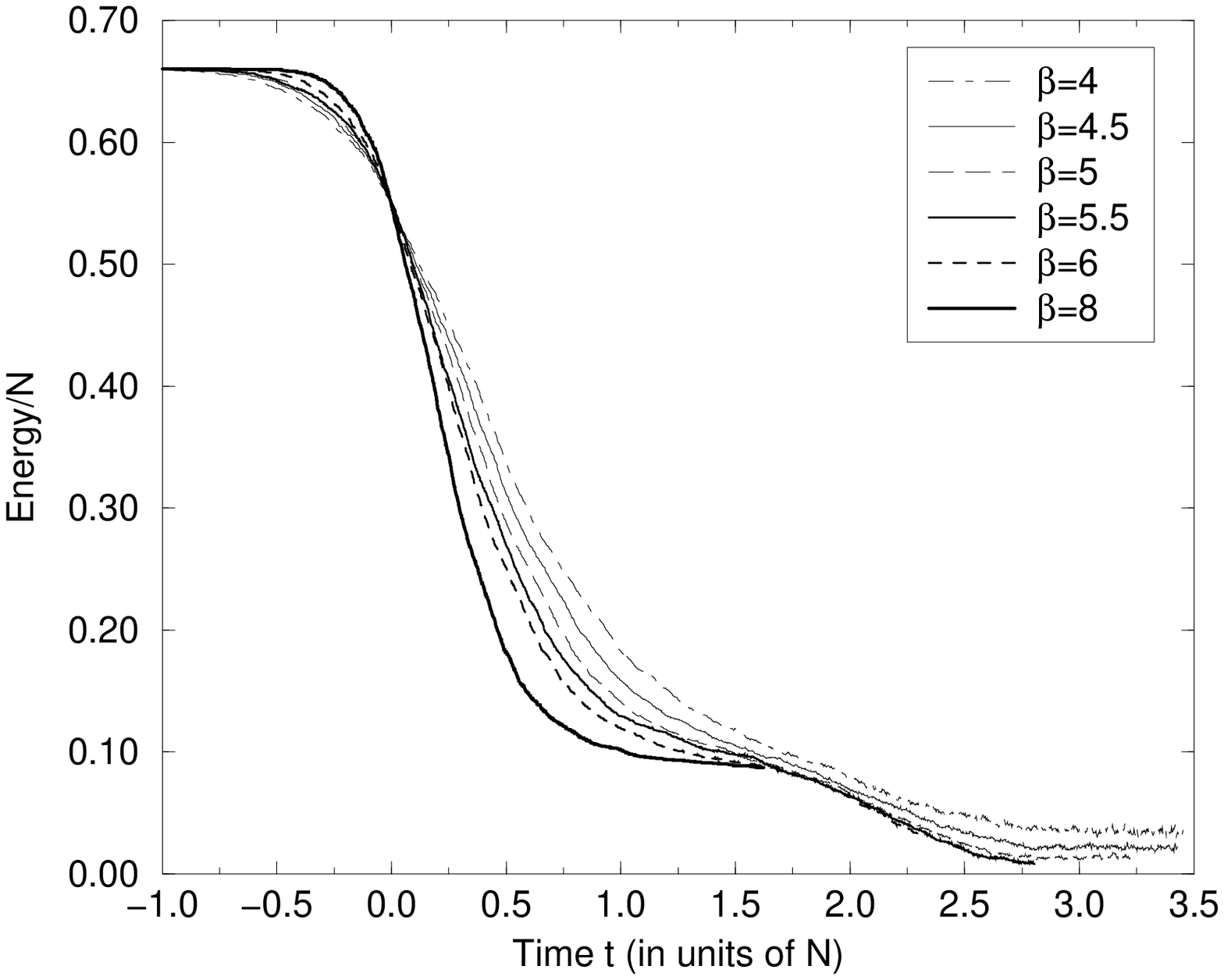}}}
\caption{\textbf{The behaviour of the energy with time.} \label{engy}}
\end{center}
\end{figure} 
\begin{figure}[t!]
\begin{center}
\subfigure[Annihilation of two dimers. This is also possible if each
spin is multiplied by -1.]{\resizebox{!}{100pt}{\includegraphics{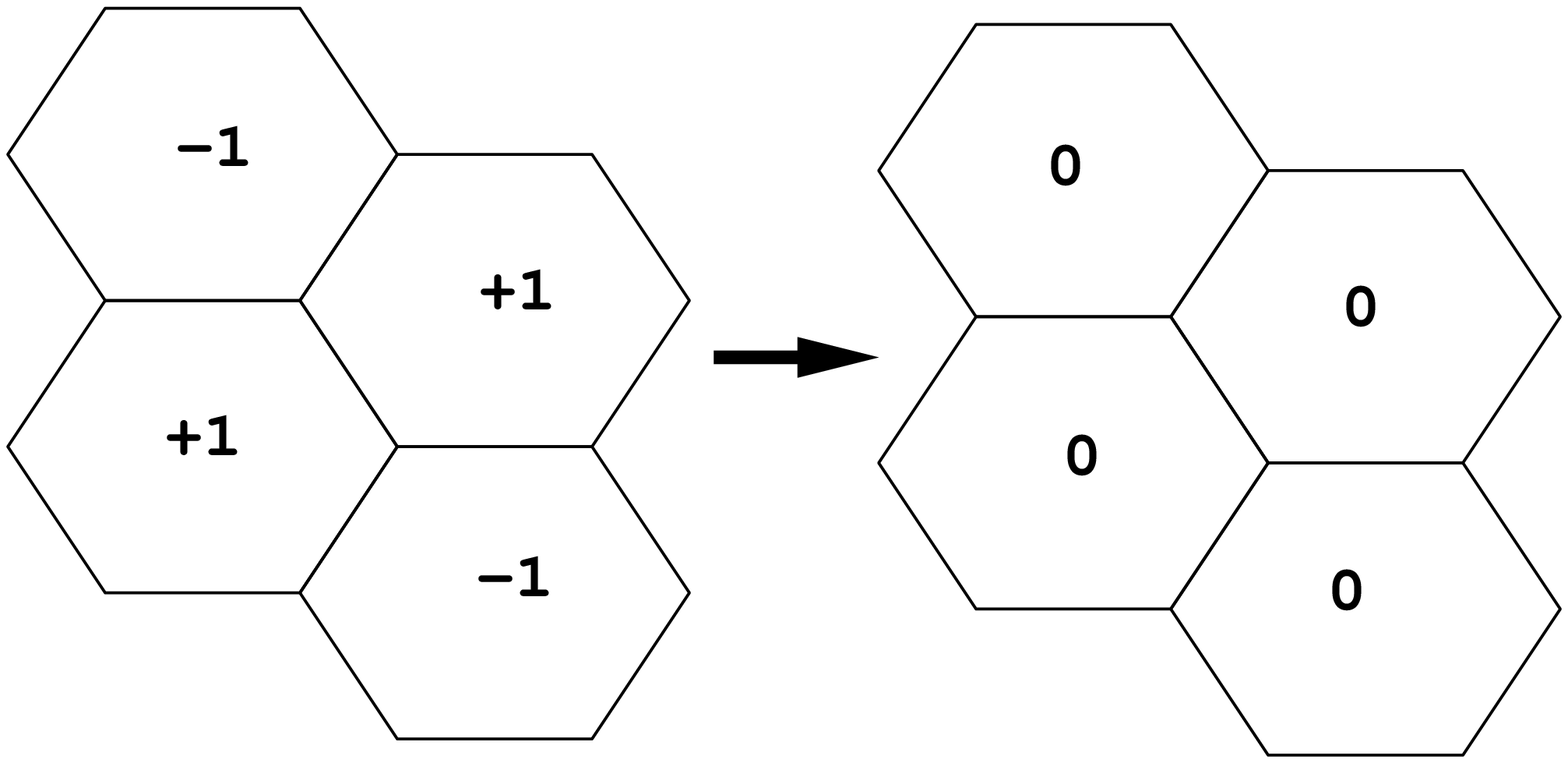}}\label{kill4}}
\hspace{2pt}
\subfigure[Annihilation of a dimer through interaction with a
defect. The defect is shifted in position.]{\resizebox{!}{100pt}{\includegraphics{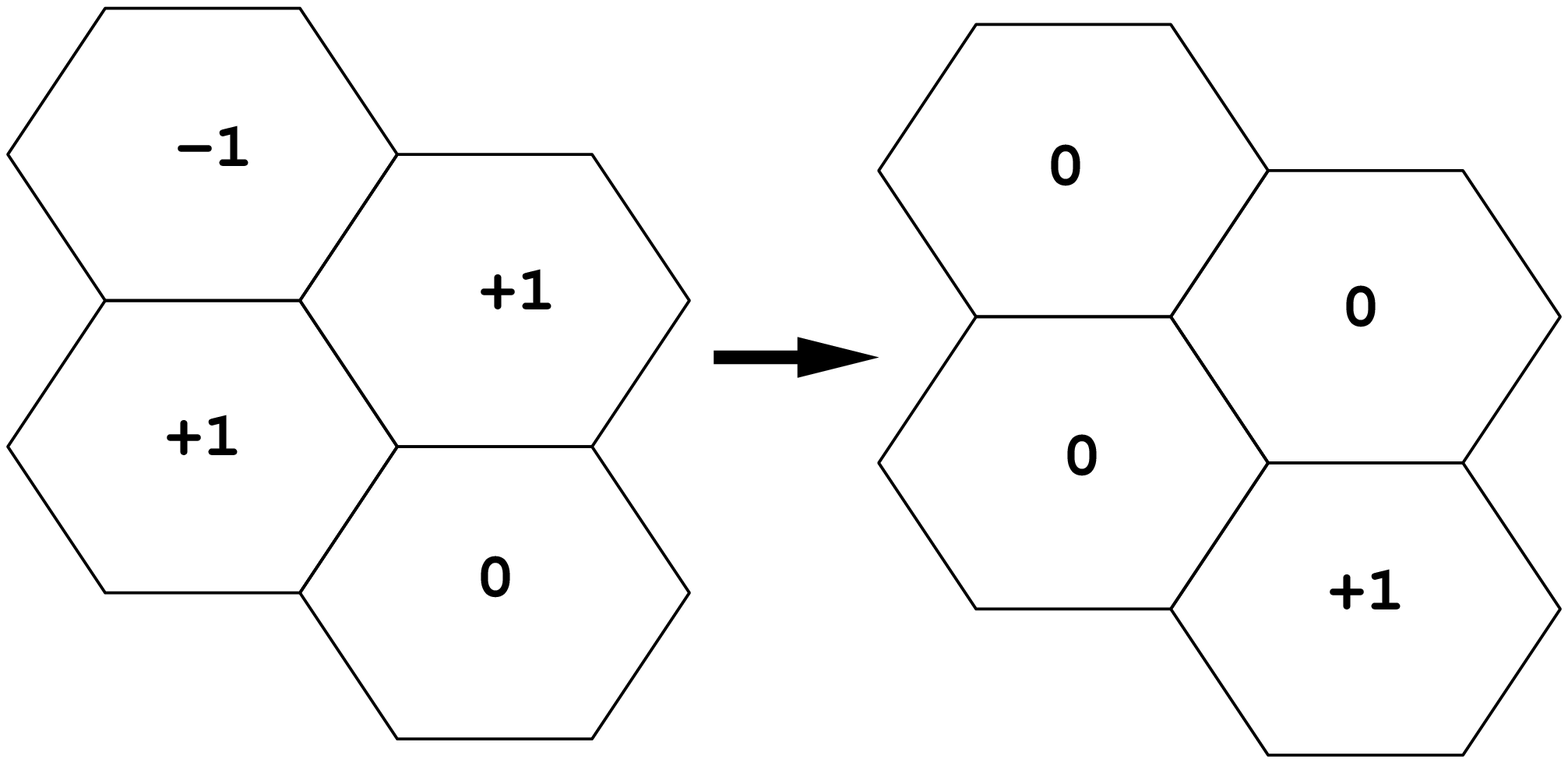}}\label{absorb}}
\subfigure[Free diffusion of a dimer in a background of zero spins.]{\resizebox{!}{100pt}{\includegraphics{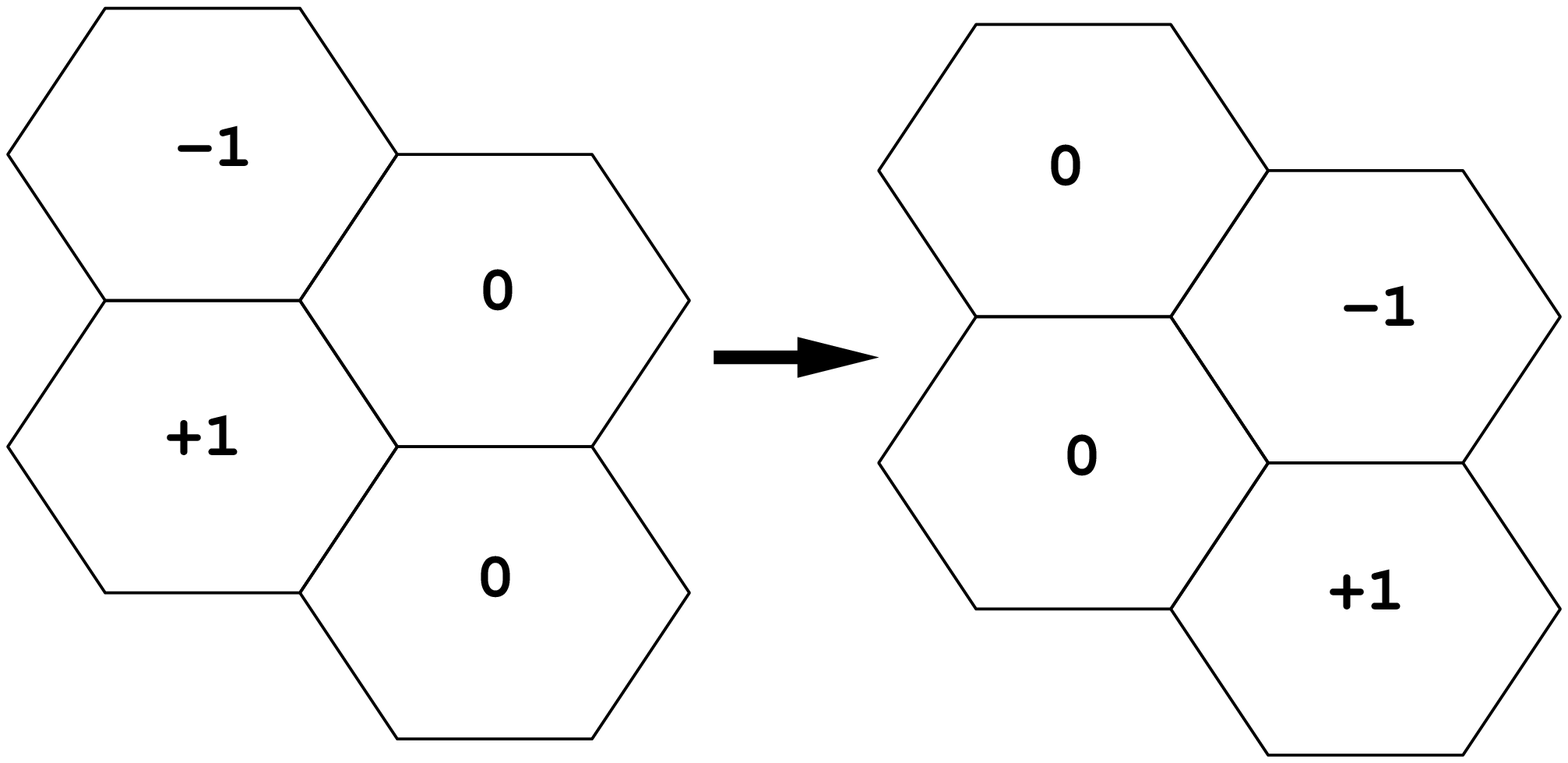}}\label{diff}}
\caption{\textbf{The dominant moves through which this model evolves.}
\label{allmoves}}
\end{center}
\end{figure} 
\begin{figure}[t!]
\begin{center}
\resizebox{!}{270pt}{\includegraphics{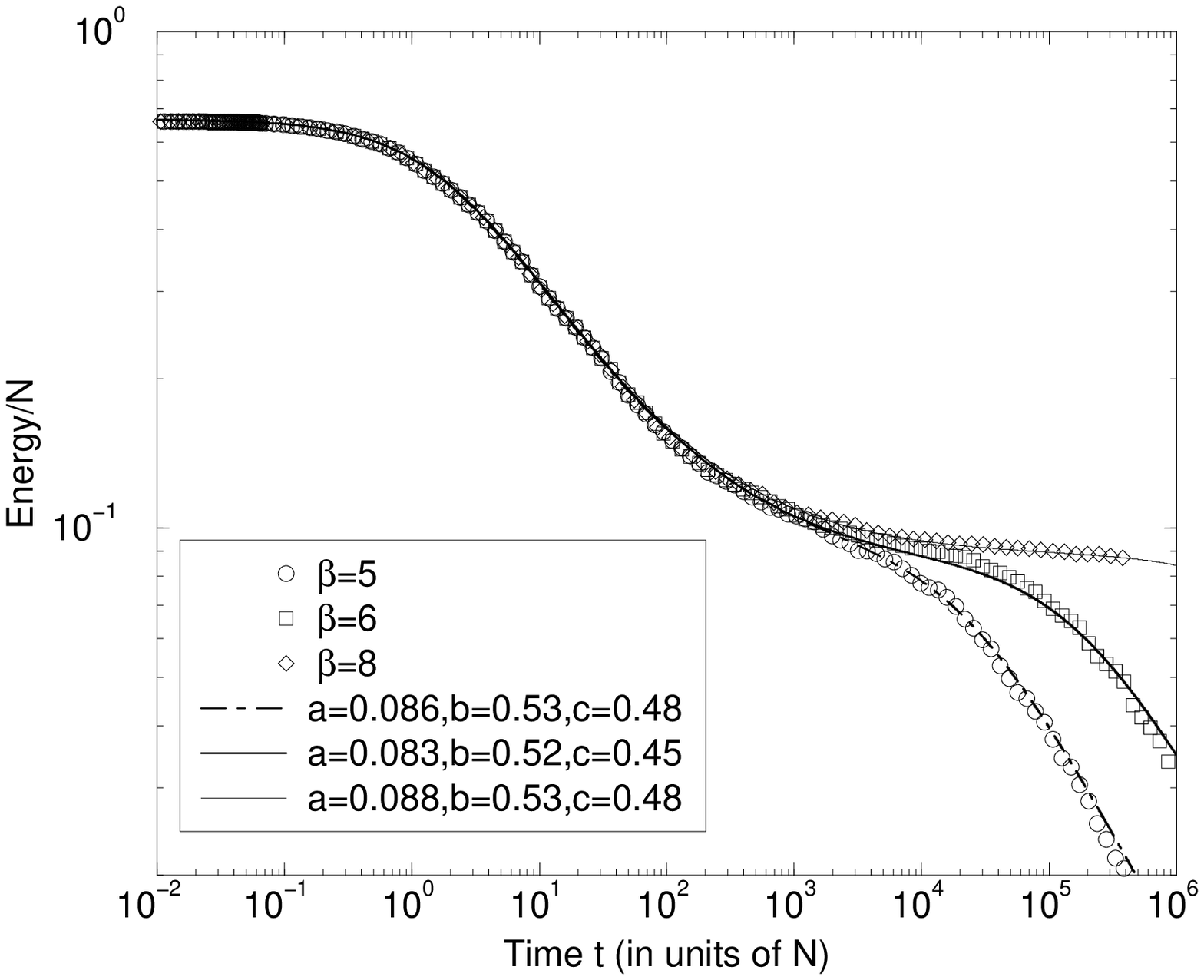}}
\caption{The energy fitted with equation (\ref{nrjfit}), using the
values shown in the
key.\label{hngy}}
\end{center}
\end{figure} 

\section{$\boldsymbol{D>0}$}
\subsection{Relaxation Dynamics}

It is a simple matter to calculate the equilibrium behaviour of
$\frac{E}{N}$, for which one finds (in units of $D=1$):
\begin{equation}
\label{mu2}
\frac{E}{N}=\frac{2\exp(-\beta)}{1+2\exp(-\beta)} \cdot
\end{equation}
One has instant access to equilibrium states, as one can randomly
place the appropriate number of $\pm1$'s throughout the system to
access a particular temperature. Using this feature, we may study the
behaviour of the energy with temperature at a number of different
cooling rates, starting from equilibrium at any chosen
temperature. Above $T=1$ the system equilibrates very rapidly at
even the fastest cooling rate, so we have chosen a starting configuration of $T=1$, and then cooling is carried out by waiting
a time $t_w=\gamma N$ at each temperature decrement of $\delta
T=0.05$. The results (averaged over 3 runs), with the equilibrium curve, are shown in Figure
\ref{hcool}; the system exhibits strong dependence of the energy upon the
cooling rate. This is characteristic of glassy systems, and
qualitatively similar to the results found for the purely topological froth.

It is instructive to study also the behaviour of the system when
subjected to a rapid quench from an infinite temperature ($\beta=0$)
configuration to a temperature at which it is allowed to evolve for a
time $t_w$. Figure \ref{hquench} shows the results from such a quench
for a range of different temperatures. For longer values of $t_w$ we
see a minimum develop - this strongly suggests activation is
present: at very low temperatures the system cannot overcome the
energy barriers and thus cannot access lower energy states. The
temperature at which this minimum occurs is dependent on the waiting
time $t_w$. At very low temperatures, even at the largest waiting
times employed ($t_w=100000N$) the system is unable to reach energies
below $\frac{E}{N} \sim 0.09$. 

One sees the significance of this value if we turn to the temporal behaviour of the energy. Figure \ref{enrgy} is
a plot of $\frac{E}{N}$ against time, quenched from a fully disordered
starting configuration corresponding to $\beta=0$ to the temperature
in question. The initial decay of the energy is fast and independent
of temperature until $\frac{E}{N} \sim 0.09$, at which point one sees the
existence of a plateau; the time spent on this plateau is clearly
dependent on temperature. Upon departure from the plateau, the energy
relaxes directly to the appropriate equilibrium value; on the graph
shown, the curves for $\beta \leq 5$ equilibrate within the time-scale
of the simulation, whereas those for $\beta >5$  do not.  This
two-time behaviour, with one time-scale temperature independent and
the other increasing with inverse temperature, can be understood as
follows. The mechanism for lowering the energy is the
annihilation of pairs of adjacent $+1$ and $-1$ spins (we
shall refer to a $+1,-1$ pair as a `dimer'). This occurs in two ways:
(i) two neighbouring conjugate dimers can destroy each other to leave
four zero spins as shown in Figure \ref{kill4}, with a reduction in energy of
4 units, or (ii) a $\pm 1$ can annihilate a dimer, thus shifting its
position and reducing the energy by 2 units as in Figure
\ref{absorb}; note that the $\pm 1$ can be part of a dimer at a
different orientation. Any such local arrangements present in the starting
configuration will be eliminated quickly without need of any thermal
excitation. Furthermore, dimers can move freely through a background of
zero spins as shown in Figure \ref{diff} until they reach a local
environment which favours annihilation, such as those previously
mentioned. This diffusion occurs on a time-scale of 2 steps per
spin, as one can alter the configuration on the left of the arrow in Figure \ref{diff} in 2 ways, one of which will be
possible and one of which will be forbidden through the
$\delta$-functions in equation (\ref{prob}). Consequently, the initial fast decay of the energy is
temperature-independent, and of a diffusive character with an
underlying time-scale of 2.  This fast, diffusive
process describes the behaviour of the energy until the plateau is
reached. To describe it further, one has to consider isolated defects
i.e. isolated spins of $\pm 1$. To remove these defects, they must be
paired up with a conjugate isolated defect to form a dimer which can
then diffuse freely as in Figure \ref{diff}, and eventually annihilate as in
Figure \ref{kill4} and \ref{absorb}. An isolated defect can move through fortuitous
collisions with existing dimers, but after the initial fast decay
these dimers become rare. Alternatively, a defect can move by creating a new
dimer (reversing the arrow in Figure \ref{absorb}), at an activation energy cost of 2 units and
with a probability that scales as $\mathrm{e}^{-2 \beta}$. As is
clear from this figure, one can interpret the resulting configuration
as a dimer-plus-defect in two ways; either of these two possible
dimers can diffuse away freely if adjacent to two zero spins, or
annihilate if adjacent to another dimer. This factor of 2 will cancel
that introduced by the diffusive time-scale of the dimers. Thus again one has a diffusive process
leading to a final reduction in energy, in this case with a time-scale of
$\mathrm{e}^{2 \beta}$ (since the time-scale for annihilation of the dimer
is negligible compared to that to produce the dimer for
$\mathrm{e}^{2\beta} \gg 1$). Thus $E(t)/N$ is expected to consist of two
diffusive processes: a fast process of time-scale 2 decaying to a
state of isolated defects, and a slow process of time-scale
$\mathrm{e}^{2 \beta}$ decaying to the equilibrium configuration. 

The plateau in $E(t)$/$N$ can be seen more clearly if the
time axis is rescaled to $T\ln t$ as in Figure \ref{lognrj}; as the
temperature is decreased ($\beta$ increased) the curves tend to a
sharp staircase form. This is reminiscent of the results found under
such rescaling for other kinetically constrained models \cite{fredrikson,
eisinger,kurchan,  sollich, RitortQ, juanpe,barratloreto}; in these other models one observes several plateaux
corresponding to several characteristic activation energies, but in
this particular case the situation is simpler as there is only one
dominant characteristic activation energy.

Both the fast dimer-dimer annihilation and the slow defect-antidefect
pairing are of the type usually designated as $A + B \to \emptyset$
\cite{toussaint,mattis,hinrich}. In the
fast process, $A$ and $B$ are dimers and `anti-dimers' i.e. a (+1,-1) dimer
annihilating with a (-1,+1) anti-dimer; for the slow process, $A$ and $B$
are isolated defects of opposite sign. In detail these diffusion processes are
more complicated than simple diffusion,  but for $A + B \to \emptyset$
processes the standard asymptotic behaviour of the density is of the
form $t^{-\frac{d}{4}}$, where $d$ is the dimensionality. Therefore we
suggest the same asymptotic $(t/\tau)^{-0.5}$ behaviour
for each process,\footnote{In fact, whilst the slow processes are
isotropic, this is not the case for the fast dimer diffusion: the
latter involves zig-zags at $30^{\circ}$ to the axis
perpendicular to the common edge between the two cells constituting
the dimer.} and fit the following form to the energy:
\begin{figure}[t!]
\begin{center}
\resizebox{!}{270pt}{\includegraphics{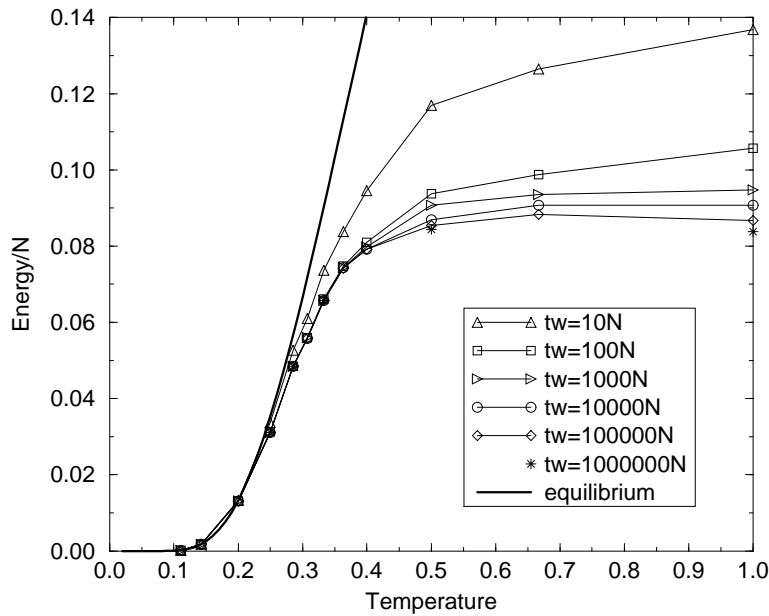}}
\caption{The behaviour of $E/N$ after a
$T=0$ quench from equilibrium. \label{heis}}
\end{center}
\end{figure} 
\begin{equation}
\label{nrjfit}
\frac{E(t)}{N}=\left( \frac{2}{3}-a \right) \left( 1+\frac{t}{2} \right) ^{-b} +
\left(a-e_{eq} \right)\left( 1+ \frac{t}{\mathrm{e}^{2\beta}}\right)
^{-c} + e_{eq}
\end{equation}
where $a$ is the plateau value, $e_{eq}$ is the energy per
spin in equilibrium, and we expect both $b$ and $c$ to be
approximately $0.5$. The results are shown in Figure \ref{hngy}; it is clear that these fits are
extremely good. In principle, the plateau value $a$ can be calculated but
here we note only that, as required,  $a$ is less than $0.25$, which is
the maximum energy for a $T=0$ frozen state. It is also less than the
value corresponding to randomly removing dimers from the initial
configuration to leave only isolated defects, which gives $E(t)$/$N
\simeq 0.2$; in fact, one finds $a \sim 0.085$ due to the effect of
singleton pairing by dimer collisions. Further investigation of the
characteristic time-scales is given below, in connection with the
correlation function.

The reader will note that the fast dimer absorption by an isolated
defect has been neglected in the above fit; this can be characterised as
type $A + C \to \emptyset + C$, and standard asymptotic behaviour of the
density for such a process is that of a stretched exponential \cite{hinrich,mattis,kayser,grassberger}. There
is also a move-set which involves two dimers interacting `off-centre',
such that one of the dimers and only one of the defects in the other
dimer are altered. This leaves either two isolated defects or a
dimer. However, in view of the excellent quality of the fit we do not
consider these explicitly at this stage.

We have also investigated the behaviour of the energy if one quenches
to $T=0$ from an initial equilibrium configuration corresponding to a finite
temperature $T_I$. This is in some sense an
investigation of the inherent states of the system
\cite{still1, still2, still3, sastrynat, sciortLJ}. At
very low temperatures, the equilibrium state for $T_I$ consists mainly of
isolated defects - thus when quenched to $T=0$ the system very quickly
reaches the inherent state. However, at higher initial temperatures this takes
an extremely long time to happen. Thus we show in Figure \ref{heis}
the energy measured at waiting time $t_w$ after a quench from the
equilibrium configuration to $T=0$. At low $T_I$ the energy
stays on the equilibrium curve, as there are no energetically
favourable moves to be made. However, for higher $T_I$ the energy tends
towards a constant value given by the plateau in
Figure \ref{engy}. Thus we see that if one quenches from an equilibrium
configuration with energy above the plateau, the lowest energy one can
reach is that of the plateau - to decrease energy further would
require activated processes, which one cannot perform at $T=0$. If one
quenches from an equilibrium start-point with energy below that of the
plateau, one cannot decrease the energy by much as there are very few
energetically favourable moves to be made, and so the curves in Figure
\ref{heis} do not
deviate far from the equilibrium curve. The temperature at which one
sees a crossover between these two types of behaviour is $T \sim 0.35$:
this is the temperature at which activated processes become
important. 

\begin{figure}\phantom{*}
\begin{center}
\subfigure[Equilibrium correlation functions $C(t)$ for (from left to right) $\beta~=~1,2,3,3.5,4,4.5,5,5.5,6$.]{\resizebox{!}{260pt}{\includegraphics{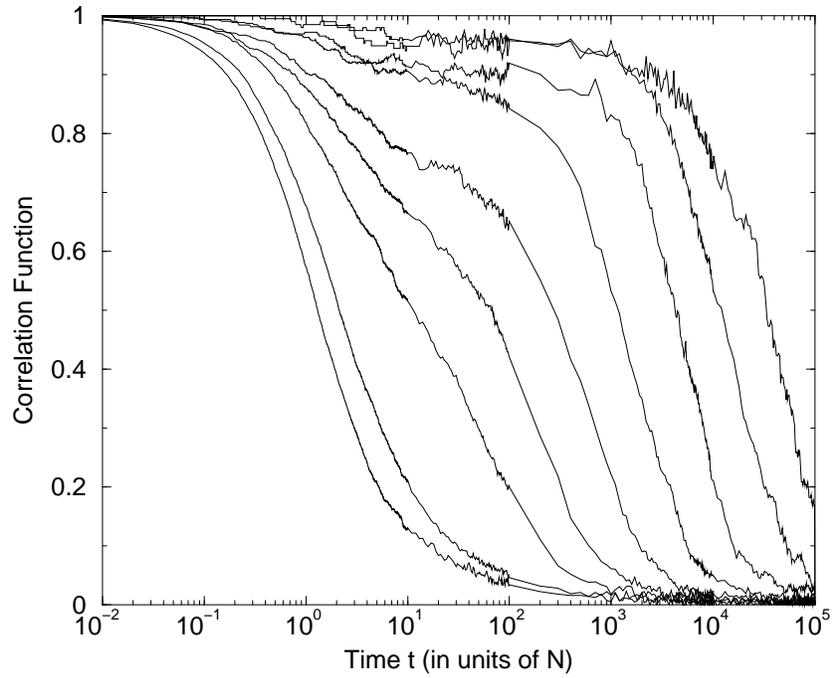}}\label{heqcor}}
\subfigure[Out of equilibrium correlation functions $C(t_w, t_w+t)$ for $\beta=6$ for (from left to right)
$t_w=10^2N, 10^3N, 10^4N$.]{\resizebox{!}{260pt}{\includegraphics{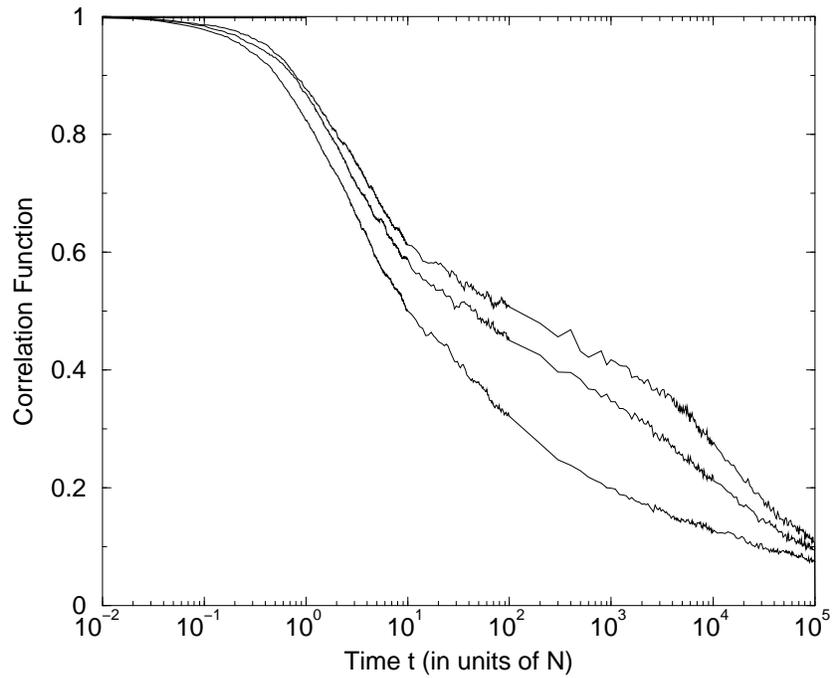}}\label{6hneq}}
\caption{\textbf{Correlation functions both in and out of equilibrium}}
\end{center}
\end{figure}

\subsection{Correlation functions and relaxation time}

We investigate the temporal correlations in the system through a two-time single-site spin correlation function of the form
\begin{equation}
\label{correl}
C(t_w,t_w +t)=\frac{\sum_{i=1}^N s_i(t_w) s_i(t_w
+t)}{\sum_{i=1}^N s_i^2(t_w)} \cdot
\end{equation}
Under equilibrium conditions this becomes a function of the relative time $t$ only.  The equilibrium
correlation functions (averaged over 5 runs) for a range of temperatures are shown in Figure
\ref{heqcor}. As $\beta$ increases, plateaux develop, revealing that
again two-step relaxation is taking place. We also show examples of typical 
out-of-equilibrium correlation functions in Figure \ref{6hneq}; one can see
aging behaviour, with the correlation function showing dependence on the
waiting time $t_w$ after a rapid quench from infinite temperature.

\begin{figure}
\begin{center}
\resizebox{!}{280pt}{\includegraphics{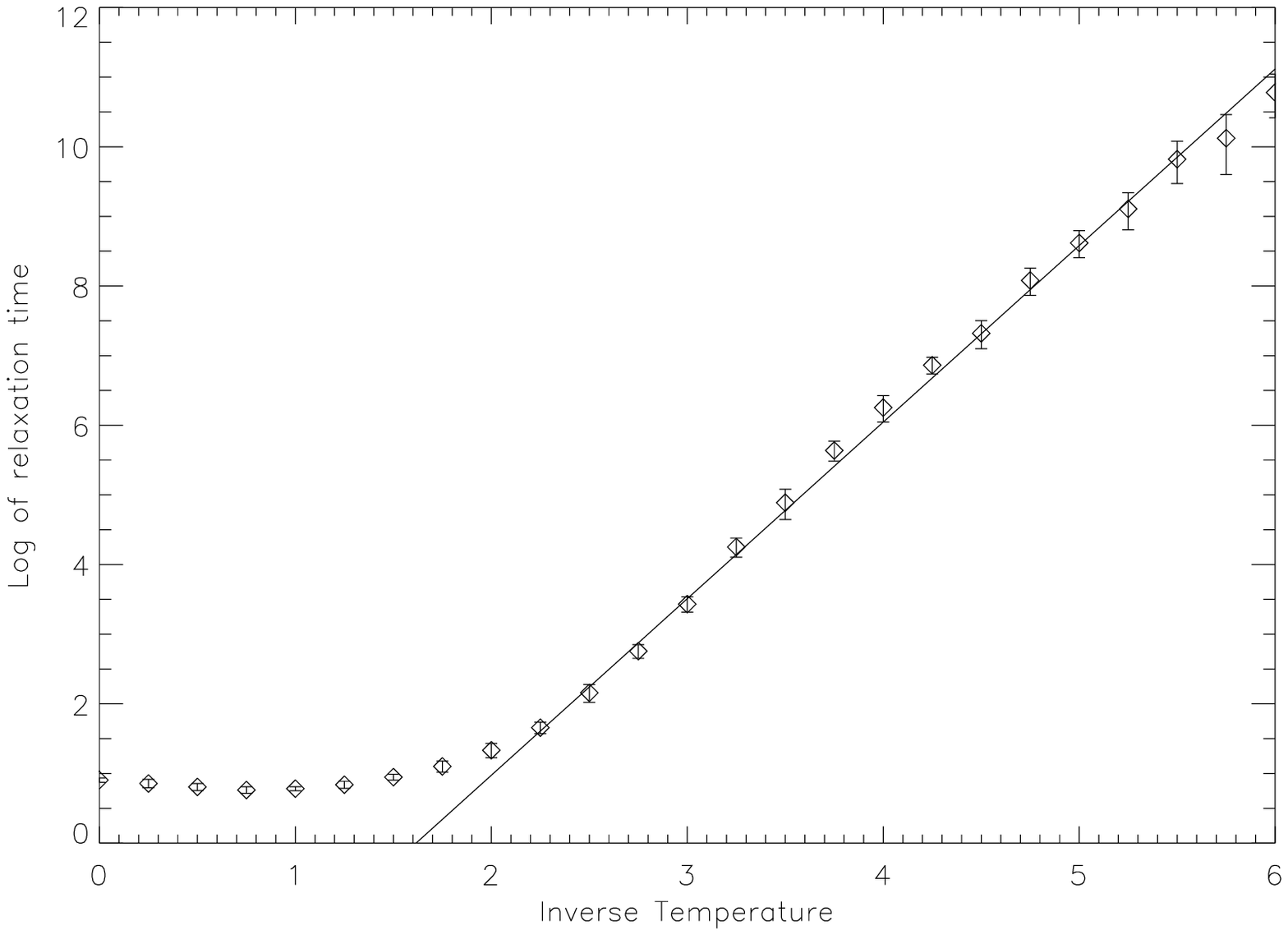}}
\caption{\textbf{The logarithm of $\boldsymbol{\tau_r}$ against
inverse temperature.} The solid line is an Arrhenius law of the form
$\tau_r=0.0166 \hspace{1pt}\mathrm{e}^{2.535/T}$.\label{rel}}
\end{center}
\end{figure}

\begin{figure}\phantom{*}
\begin{center}
\resizebox{!}{100pt}{\includegraphics{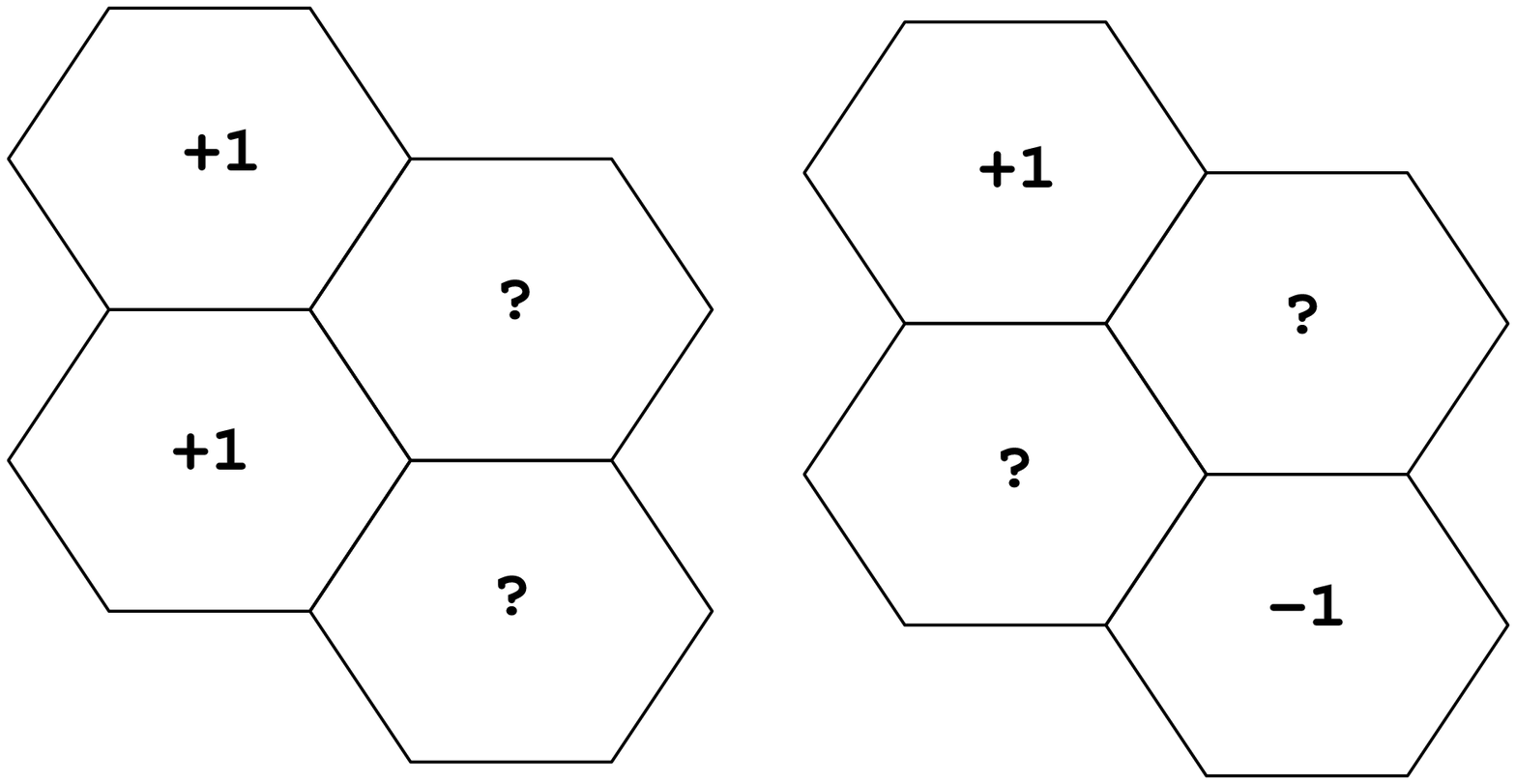}}
\caption{\textbf{`Stuck' configurations.} On the left, one sees that
two spins of $+1$ adjacent to each other are unable to be flipped
within the isolated cluster,
regardless of what values the other two spins take (and similarly for
two adjacent spins of $-1$). On the right, we see that two
next-to-nearest neighbours of opposite spin cannot be flipped,
regardless of the values of the other two spins. A favourable
configuration external to the cluster can provide an opportunity to
`unstick' these cells.
\label{stuck}}
\end{center}
\end{figure}

Following the procedure in  \cite{kobreview,paper1}, we may define a
relaxation time $\tau_r$ as the time at which the equilibrium 
correlation function decays to $\mathrm{e}^{-1}$. This is plotted on
Figure \ref{rel} against inverse temperature; the data can be
reasonably fitted by an Arrhenius curve of the form
\begin{equation}
\label{arrhen}
\tau_r=A \hspace{2pt} \mathrm{e}^{B/T}
\end{equation}
where $A,B$ are constants. The solid line superimposed on
Figure \ref{rel} corresponds to $A=0.0166, B=2.535$. This
indicates that this model displays strong glassy behaviour, in
agreement with the results from the topological model, where an offset
Arrhenius law fitted the data considerably better than either a power
law, or a Vogel-Fulcher law \cite{paper1}. The
anomalous upturn in the curve at very low $\beta$ is due to
the restriction that the spins may only take the values $\pm 1$ or
$0$; when the density of non-zero spins is very high, it becomes
likely that some defects will sit next to each other, in locally
`stuck' configurations which slow the decay of the correlation
function (see Figure \ref{stuck}).

However, if we look closely at Figure \ref{rel}, we see that at high
inverse temperature the data seems to be drifting below the curve. If we
consider more carefully the expected form of $C(t)$ it becomes clear
that the above definition of $\tau_r$ is not the most appropriate one,
since again, as in the toplogical model, there are two decay processes. One
can understand the origin of the plateaux by considering further the dominant
processes involved in evolution of the system, which are analogous to
those in the topological model \cite{paper1}. In equilibrium, at low temperatures
(high $\beta$) there are few defects present in the
system. The initial fast decay from $C(t=0)=1$ is due to dimers
diffusing freely through the system, and thus moving the system away from the
starting configuration. Isolated defects, however, need to either
absorb or create a dimer in order to change position; this happens on
a much longer time-scale than diffusion of the dimers. Therefore again
we have two
time-scales in the model - fast dynamics due to diffusion of the
dimers, and slow dynamics due to movement of the isolated or stuck
defects through absorption/creation of dimers. As noted earlier, the former is temperature-independent as dimer
diffusion costs no energy; however, the position of the plateau and the
subsequent departure from it is dependent on temperature. In
particular, the depth of the drop to the plateau from the initial
$C(t=0)=1$ is determined by the
equilibrium concentration of spins of $\pm 1$ in local configurations that
are free to move with no change to the energy (these local configurations
include the dimers). One can
therefore suggest that $C(t)$ might be a sum of two functions in the
following way:
\begin{equation}
\label{correltry}
C(t) = \alpha \hspace{2pt}f(t, \tau_1) + (1- \alpha)\hspace{2pt} g(t, \tau_2)
\end{equation}
where $\alpha$ and $\tau_2$ are functions of T, but $\tau_1$ is a
constant, and $\alpha$ is the height at which one would expect to find
a plateau. $f$ and $g$ are functions to be determined, but they must
be monotonically decreasing functions of $t$, satisfying $f(0,\tau_1)=g(0,\tau_2)=1$ and $f(\infty,\tau_1)=g(\infty,\tau_2)=0$.

\begin{figure}\phantom{-}
\begin{center}
\resizebox{!}{250pt}{\includegraphics{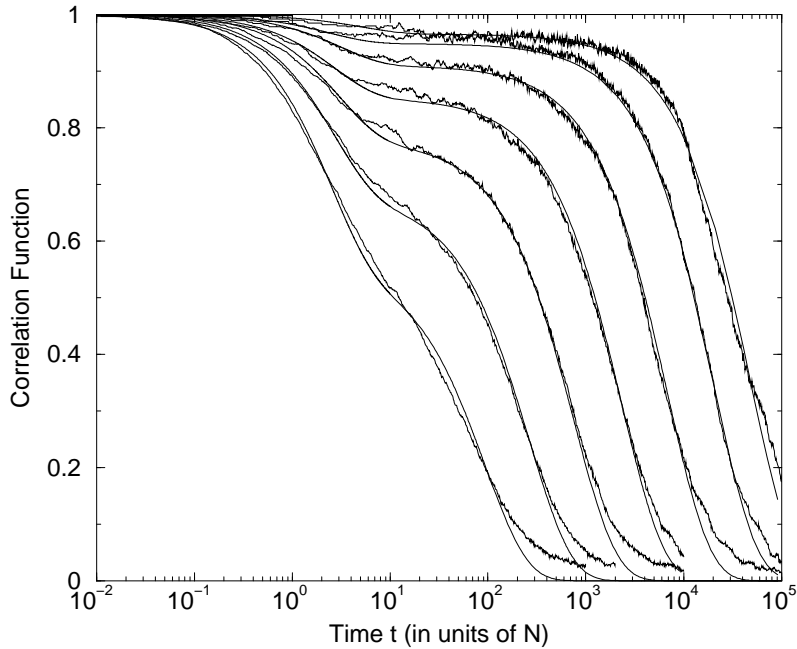}}
\caption{Correlation functions for, from left to right,
$\beta=3,3.5,4,4.5,5,5.5,6$. The solid lines superimposed are the best fits
of the form $C(t) = \alpha \hspace{2pt} \mathrm{e} ^{-t/\tau_1} + (1- \alpha)
\hspace{2pt} \mathrm{e}^{-t/\tau_2}$.\label{corfit}}
\end{center}
\end{figure}

\begin{figure}
\subfigure[$\tau_1$ against inverse temperature.]{\label{hetau1}\resizebox{!}{200pt}{\includegraphics{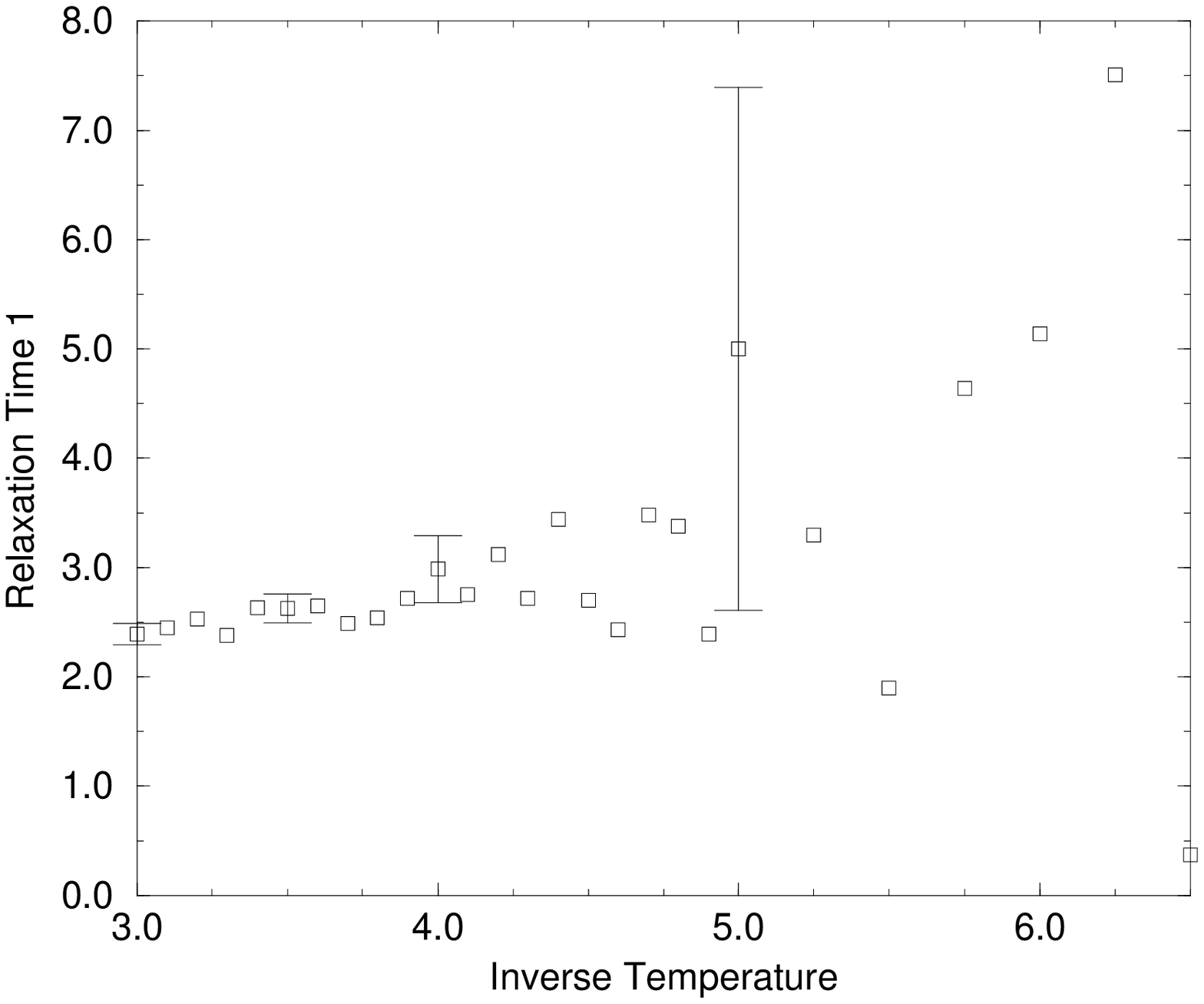}}}
\subfigure[$\tau_2$ against inverse temperature; the superimposed line is
$y=A\mathrm{e}^{B\beta}$, where $B=2.12$. ]{\label{hetau2}\resizebox{!}{200pt}{\includegraphics{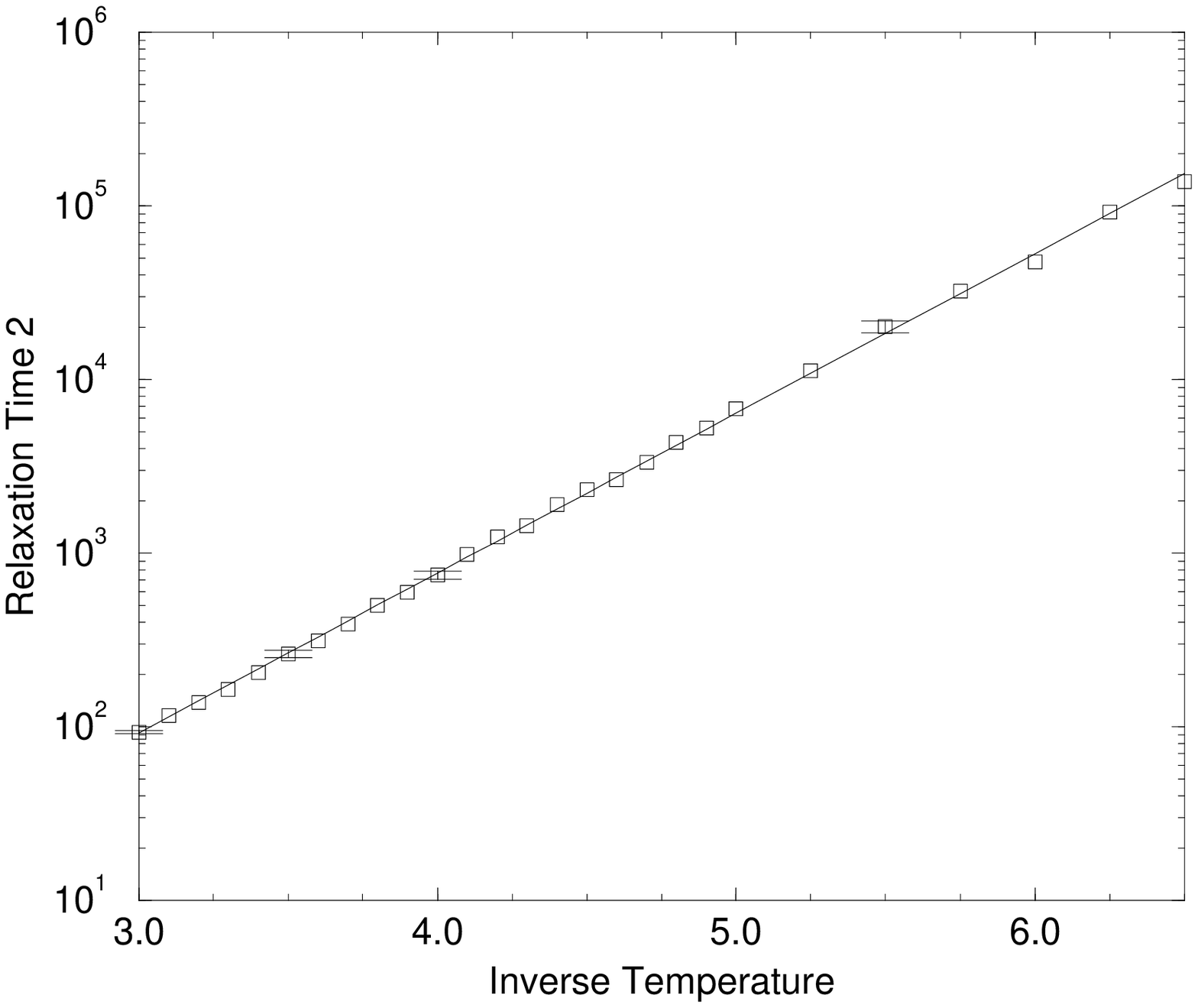}}}
\begin{center}
\subfigure[$\alpha$ against inverse
temperature;
the superimposed curve is $y=\frac{12
\mathrm{e}^{-\beta}}{(1+2\mathrm{e}^{-\beta})^3}$]{\label{heplat}\resizebox{!}{200pt}{\includegraphics{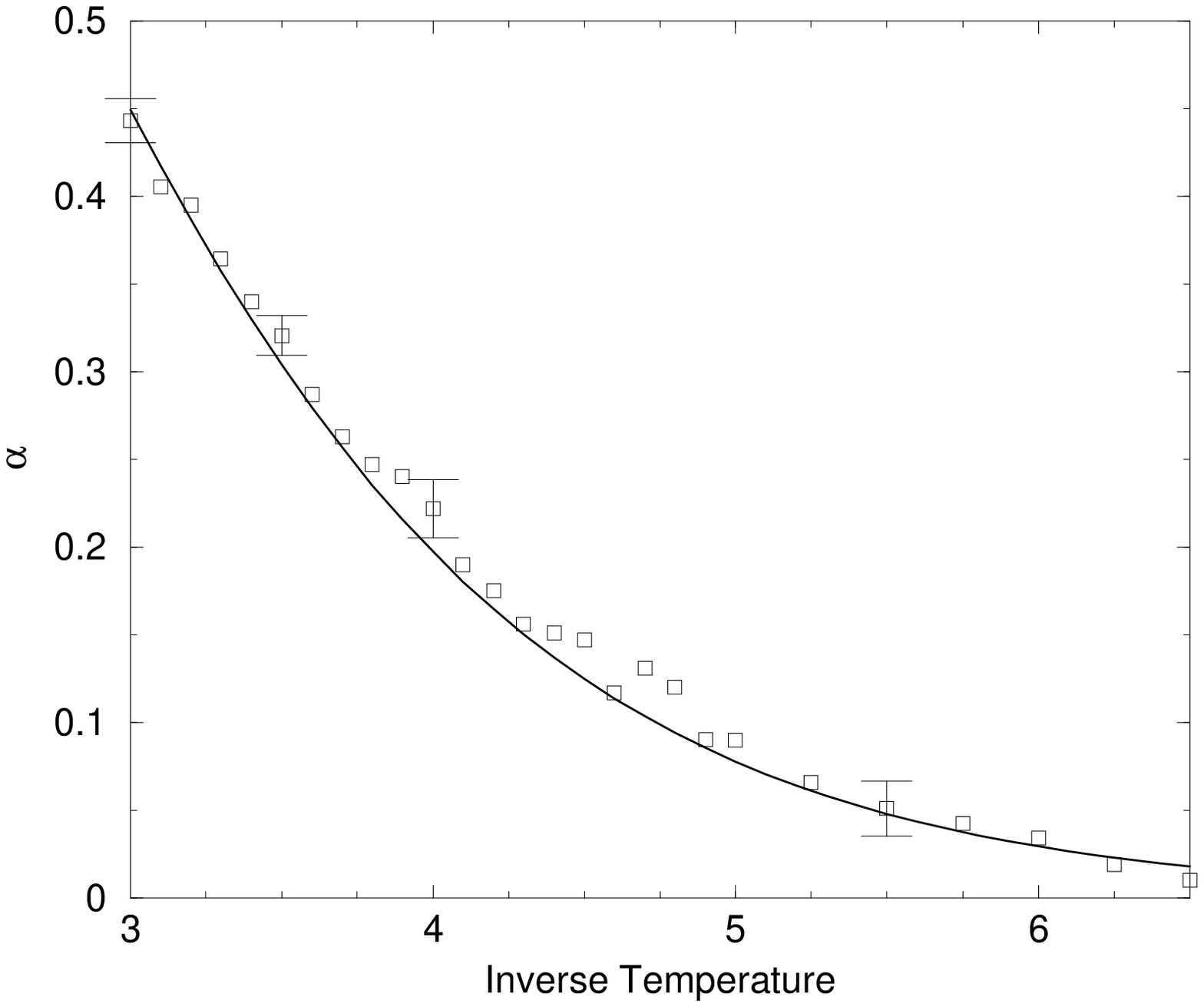}}}
\end{center}
\caption{\textbf{The behaviour of the fitting parameters
$\boldsymbol{\tau_1,\tau_2}$ and $\boldsymbol{\alpha}$ with temperature};
obtained by fitting $C(t) = \alpha \hspace{2pt} \mathrm{e} ^{-t/\tau_1} + (1-\alpha)
\hspace{2pt} \mathrm{e}^{-t/\tau_2}$. \label{hefits}}
\end{figure}

We tried to fit equation (\ref{correltry}) to the data using
exponentials for both $f$ and $g$ such that $C(t)$ is of the following form:
\begin{equation}
\label{correlfit}
C(t) = \alpha \hspace{2pt} \mathrm{e} ^{-t/\tau_1} + (1- \alpha)
\hspace{2pt} \mathrm{e} ^{-t/\tau_2} .
\end{equation}
The results are shown in Figure \ref{corfit}; in order to fit more
accurately, the system size has been increased to $N=160000$. The fit is extremely
good at high values of $\beta$ (low temperature), although at lower values there is some
deviation. We also note that the fitted form drifts below the data at very
low values of $C(t)$; there may be some correction to this form which
we have not taken into account. We do not claim that this form is
exactly correct; nevertheless, it is a useful approximation that may
allow us to separate out the two time-scales. Previously we stated that we expect $\tau_1$ to be independent of
temperature - Figure \ref{hetau1} shows that this seems to be the case
at low $\beta$, although as $\beta$ increases it becomes harder and
harder to fit $\tau_1$ accurately due to the extremely high
position of the plateau. We have shown the error bars on a few of the
points to give some idea of the difficulty in accurately performing
this fit at high $\beta$; one can see that it is impossible to say
anything sensible about the functional form of $\tau_1(\beta)$ for
$\beta>4.5$. However, the naive theory as discussed earlier in
connection with $E(t)$ gives $\tau_1=2$ and the data is in general accord.

We turn now to $\tau_2$ and consider the dominant processes involved
in relaxation in the $\beta$-relaxation regime. These are, as
previously mentioned, absorption and creation of dimers. Creating a
dimer costs 2 units of energy. Each dimer will rapidly diffuse freely
through the system until it is absorbed by a defect; this happens
quickly as it is an energetically favourable process. Therefore we
have energy barriers of 2 and probabilistic barriers of $2\beta$ in this
regime. Absorption of pre-existing dimers also has a characteristic
time-scale of $\sim \mathrm{e}^{2\beta}$ as the fraction of cells
occupied by dimers scales as $\mathrm{e}^{-2\beta}$. This is
reflected in the behaviour of the second time-scale $\tau_2$ as shown
in Figure \ref{hetau2}: $\tau_2$ exhibits the  Arrhenius behaviour of $\tau_2 = A \mathrm{e}^{B \beta}$,
with $B=2.12$. We have shown error bars on a
few of the points; these tend to suggest that the value of $B$ is not
exactly 2; however, it is encouraging for the value to be so close
given that this is a very crude theory.

The plateau parameter $\alpha$, is somewhat more complicated. As noted
earlier, the initial fast
decay of the correlation function is dominated by all the local
configurations that can move freely, without any energy costs. Thus
we expect the plateau height to occur roughly at a value of 1 minus the
fraction of total spins which can move freely in equilibrium
conditions. This fraction contains the dimers, but also certain configurations of like-pairs
(i.e. $+1,+1$ and $-1,-1$) which can oscillate, as shown in Figure
\ref{oscill}, although they cannot delocalise without interacting with
dimers. In equilibrium the probability that a given  defect is part of a
$+1,-1$ dimer is:
\begin{figure}[t!]
\begin{center}
\resizebox{!}{100pt}{\includegraphics{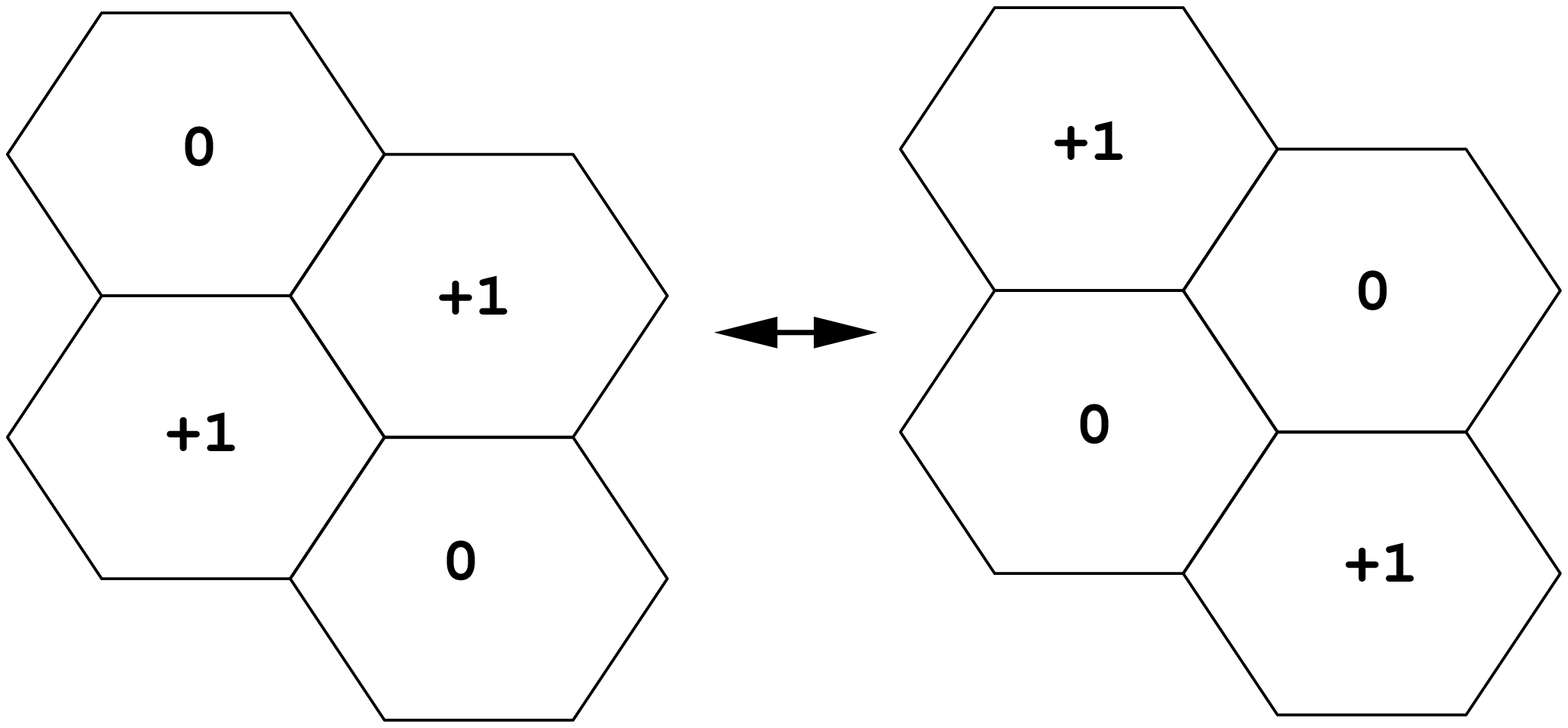}}
\caption{\textbf{Zero-energy moves for oscillating $+1$ like-pairs.} An identical
configuration exists for $-1$ like-pairs.} 
\label{oscill}
\end{center}
\end{figure} 
\begin{equation}
\label{dimerprob}
6 p(1) p(0)^2 = \frac{6\mathrm{e}^{-\beta}}{(1+2\mathrm{e}^{-\beta})^3}
\end{equation}
where $p(s)$ is the probability of a spin of value
$s$ (the factor 6 comes from the fact the second defect may be situated
on any of the 6 nearest neighbours to the original defect). The
probability that any given defect is part of an oscillating like-pair
is given by $12 p(1) p(0)^2$ (the factor 12 is because there are 12
positions at which the second defect may be situated to make up an
oscillating like-pair). However, subsequent to $t=0$ one supposes that
at any instant in time half of the oscillating like-pairs will be in
exactly the same position as at time $t=0$, and the other half will be in
the alternative position. All other processes which cost no energy
involve at least 3 defects, and thus are suppressed in comparison by
factors of $\mathrm{e}^{\beta}$ or $\mathrm{e}^{2\beta}$. As we are
investigating the region of $\beta>2.5$, we neglect those, and simply suggest:
\begin{equation}
\label{guessa}
\alpha(\beta)=\frac{12\mathrm{e}^{-\beta}}{(1+2\mathrm{e}^{-\beta})^3}
\end{equation}
Figure \ref{heplat} shows this predicted curve of $\alpha$ against the simulation
results for fitting with a sum of two exponentials, from which good
agreement can be seen. Therefore for all three parameters ($\tau_1,
\tau_2, \alpha$) the results of the simulations provide some support to the
theory of the dominant processes involved in the evolution of this system.

Note that the relaxation behaviour of the equilibrium
correlation function $C(t)$ differs from that of $E(t)$ in two
important respects. One is that for $C(t)$ the system is always in
macroscopic equilibrium so that the macroscopic distribution of
non-zero spins fluctuates around a constant value throughout the
dynamics, and all the observed results are due to re-arrangements of
the location of these non-zero spins i.e. annihilations are balanced by
creations on a macroscopic level. The second is that one has merely to
move a non-zero spin in order to affect $C(t)$, whilst for $E(t)$ to
decay from the infinite temperature starting configuration one
must actually annihilate $\pm 1$ spins overall. 

\subsection{Response and overlap functions}
We continue to show that this simple spin model behaves in the same
fashion as the topological froth by studying response functions in
relation to the fluctuation-dissipation ratio. Again we concentrate on
the single-site (averaged) case, for which we require the linear
response at a site to an infinitesimal perturbation field at the same
site. The
system is quenched from $\beta=1$ to the temperature required, and
then allowed to evolve at that temperature until time $t_w$ when a
field of magnitude $h$ and random sign $\epsilon_i = \pm 1$ is applied. Therefore
the perturbation to the energy introduced by the field/charge interaction is:
\begin{equation}
\label{pert}
\Delta E(t) = h \sum_{i=1}^N \epsilon_i s_i(t) \hspace{1pt}\theta (t-t_w) 
\end{equation}
where $\theta (t-t_w)$ is the Heaviside function: $\theta (t-t_w)=1
\hspace{2pt} \forall\hspace{2pt} t\geq t_w$; 0 otherwise. $h$ is a carefully chosen compromise which gives both linear response and a
reasonable signal-to-noise ratio. The quantity which one measures is
then the linear response function $G(t_w, t_w + t)$:
\begin{equation}
\label{response}
G(t_w, t_w + t)=\frac{\sum_{i=1}^N \epsilon_i s_i(t_w + t)}{h \sum_{i=1}^N s_i^2(t_w)}
\end{equation}
One expects a parametric plot of $-T \hspace{2pt}G(t_w, t_w + t)$ against $C(t_w,
t_w + t)$ to have a slope of -1 where the equilibrium fluctuation-dissipation
ratio is upheld. Breaking of this conventional equilibrium ratio is
characteristic of aging in glasses \cite{bouchcugreview, parisifdr};
the form of the slope when it is broken provides some information about the nature of the system. 

\begin{figure}[t!]
\subfigure[$\beta=4$. From lower curve to upper curve, $t_w=10N,
10^2N, 10^3N$. The straight line of slope -1 is shown for comparison.]{\label{4resp}\resizebox{!}{170pt}{\includegraphics{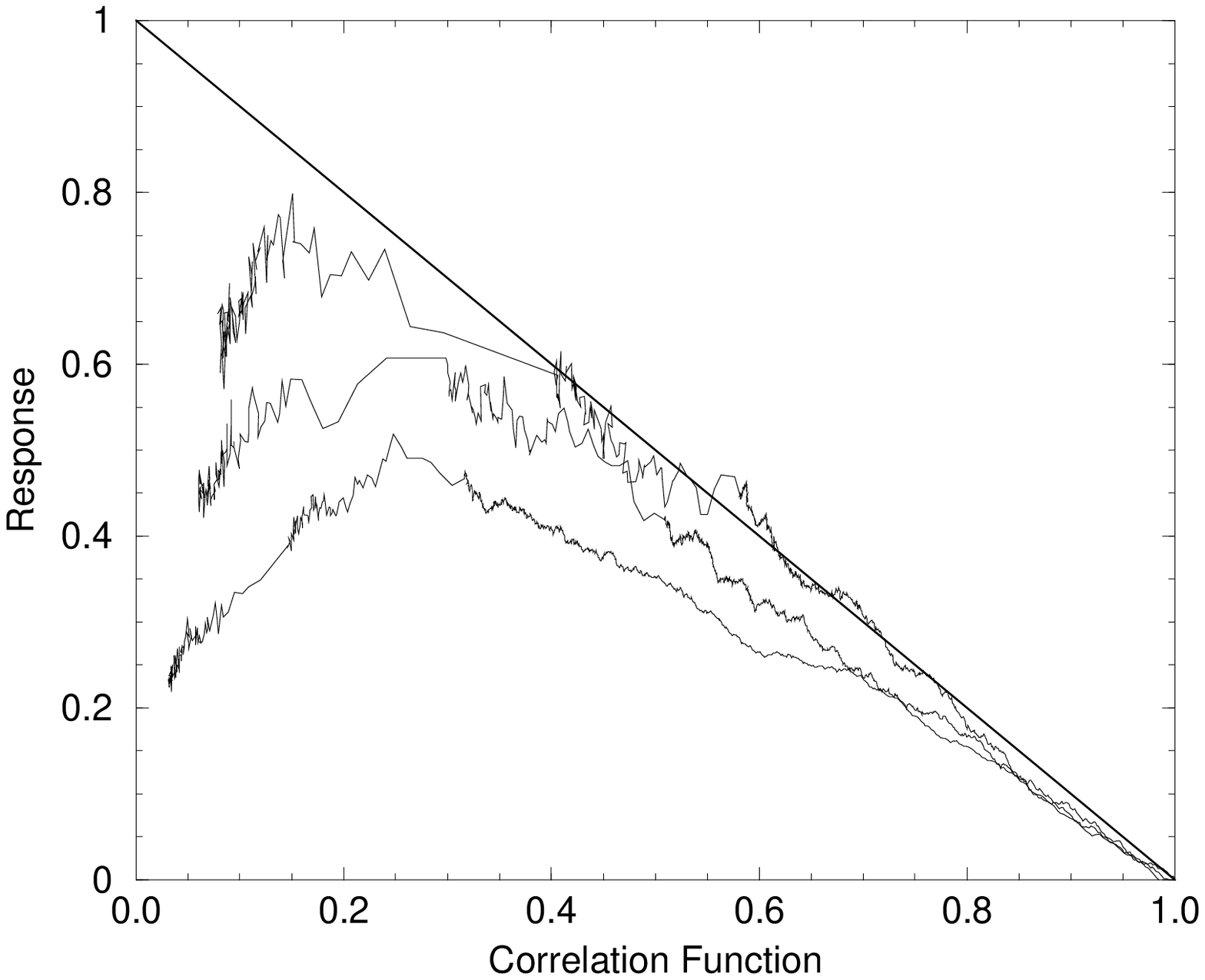}}}
\subfigure[$\beta=5$. From lower curve to upper curve, $t_w=10N,
10^2N, 10^3N$ and $10^4N$. The straight line of slope -1 is shown for comparison.]{\label{5resp}\resizebox{!}{170pt}{\includegraphics{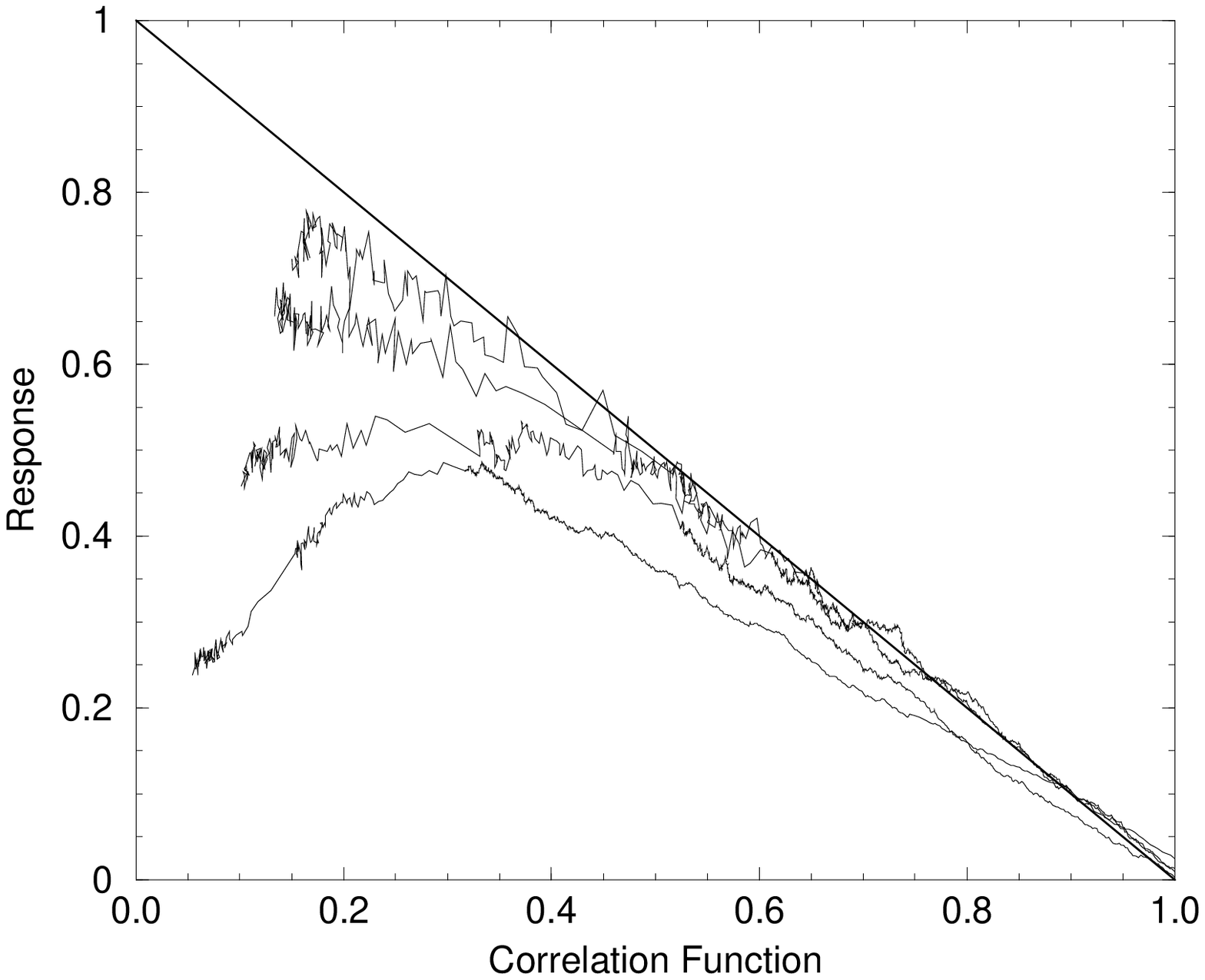}}}
\begin{center}
\caption{\textbf{Parametric plots of the response function against the
correlation function} (each averaged over
10 realisations of the charge distribution).\label{resp}}
\end{center}
\end{figure}

Figure \ref{resp} shows the results at various temperatures: these
show the same features as the topological froth, namely a
breakdown of the fluctuation-dissipation relation when $t_w$ is too
short for equilibration to have occurred (for $t_w \to \infty$ the
system is already in equilibrium when the perturbation is introduced
and FDT holds) and a reduction in the
magnitude and an eventual change in sign of the slope as $t$
increases and the correlation function decreases. The non-monotonicity
is a consequence of the existence of an absorbing equilibrium
state: competition exists between the field, which encourages the
non-zero spins to settle on energetically favourable sites and thus increases
the response, and the natural relaxation to equilibrium, which removes
non-zero spins altogether thus reducing the response (one should recall that
spins of value $0$ make no contribution whatsoever to the
response). At time $t_w$ (that is for $t=0$) when the field is switched on, the response
is on average zero. If $t_w$ is short enough that the fast processes
remain dominant (i.e. $E(t_w)$ has not yet reached the intermediate
plateau) then these energetically favourable fast moves are quickly
carried out. This simultaneously removes many non-zero spins, and also
settles many non-zero spins on energetically favourable sites; the
former process does not on average decrease the response since at
$t_w$ it is zero on average anyway, and the latter process increases
the response. Once the dimer concentration has relaxed to the
equilibrium level (i.e. after the onset of the plateau in $E(t_w + t)$) the
response increases more slowly as the evolution of the system is
dominated by the movement of isolated defects, which occurs on
much longer time-scale. The applied field causes these isolated
defects to tend to settle on
energetically favourable sites; however, as time passes they will be
removed from the system until the equilibrium concentration is reached. Therefore one expects the response to peak
before settling at a finite value once equilibrium has been reached
(the final value reached is dependent on the waiting time $t_w$ through the
normalisation of both the response and the correlation functions). 

In order to distinguish between different types of aging, Barrat,
Burioni and Mezard suggested the study of the overlap $Q_{t_w}(t)$ between two
replicas \cite{MezardQ}. These replicas are identical configurations
at time $t_w$, but are subsequently evolved according to
different stochastic thermal noise (but of the same characteristic temperature),
with $Q_{t_w}(t)$ of the form:
\begin{equation}
\label{overlap}
Q_{t_w}(t)=\frac{\sum_{i=1}^N \sigma_i^1(t_w + t) \sigma_i^2(t_w + t)}{\sum_{i=1}^N \big{(}\sigma_i^1(t_w)\big{)}^2}
\end{equation}
where $\sigma$ is some order parameter of the system and the superscripts 1,2 refer to replicas 1,2. 
 Barrat et al classify systems as either Type I or Type II models: for
 the former, the appropriate $Q_{t_w}(t)$ (normalised to 1 at time $t=0$)
decays to a finite, non-zero value in the double limit $\lim_{t_w \to
 \infty} \lim_{t \to \infty} Q_{t_w}(t)$; this class includes models which
are dominated by coarsening (for a review of coarsening see \cite{bray}). For the class of Type II, $Q_{t_w}(t)$ decays to zero in this limit; this class includes glassy systems
\cite{RitortQ}. In equilibrium (i.e. $t_w$ greater than the
 equilibration time of the system) there is no $t_w$ dependence so
 $Q_{t_w}(t)= Q(t)$; one also finds that  $Q(t)=C(2t)$ (for details see
 \cite{MezardQ}).

\begin{figure}[t]
\begin{center}
\resizebox{!}{280pt}{\includegraphics{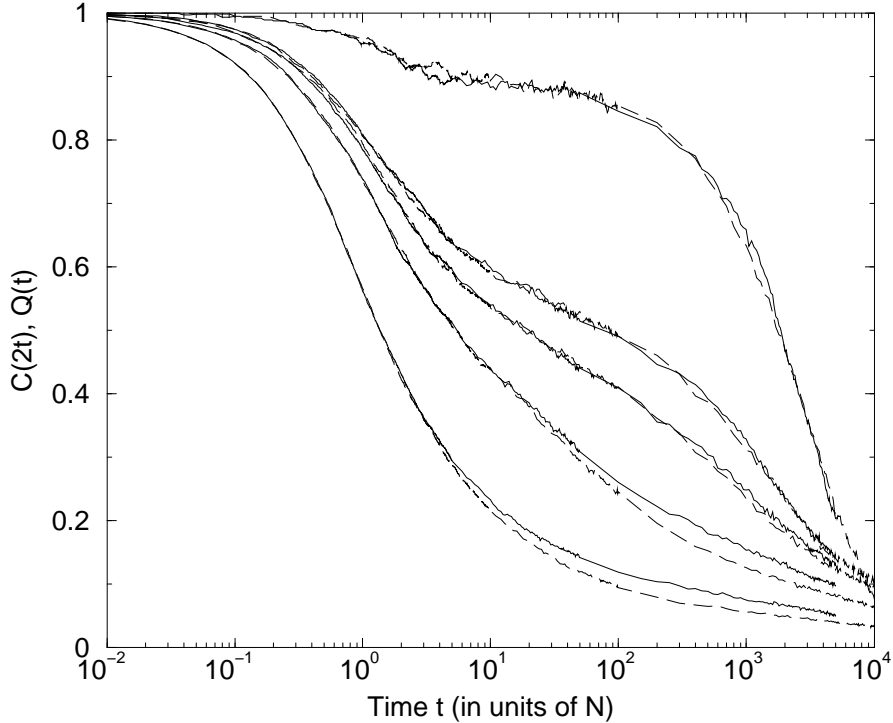}}
\caption{\textbf{The behaviour of $\boldsymbol{C(t_w, t_w+2t)}$, $\boldsymbol{Q_{t_w}(t)}$
with time for $\boldsymbol{\beta}$=5.} $Q_{t_w}(t)$ is given by the dashed curves and
$C(t_w, t_w+2t)$ by the solid curves. From lower curve pair to upper
curve pair, $t_w=10N, 10^2N, 10^3N,
10^4N$. The uppermost pair is that of equilibrium, with
$Q(t)=C(2t)$.\label{5hqct}}
\end{center}
\end{figure}

For this model, the overlap we use is as given in equation
(\ref{overlap}) with $\sigma_i = s_i$. We start from non-equilibrium
conditions, where the system is quenched at $t=0$ from $\beta=1$ to
the temperature in question, and then allowed to run at that
  temperature until time $t_w$ when measurements commence. In Figure
\ref{5hqct} we show the overlap and $C(t_w, t_w + 2t)$ against time for
$\beta=5$; initially Q is almost identical to $C(t_w,t_w + 2t)$, but it drops
below $C(t_w, t_w + 2t)$ at longer times. For larger values of $t_w$ this takes
longer to happen since these systems start off closer to
equilibrium. We also show the equilibrium curves, for which
$Q(t)=C(2t)$. 

Given that $Q(t)=C(2t)$ in equilibrium, we can check our proposed
functional form for the equilibrium correlation function through a
parametric plot of the overlap against $C(t)$. If $C(t)$ is indeed of the form given in
equation (\ref{correlfit}) then one expects:
\begin{equation}
\label{Qeq}
Q(t)=C(2t)=\alpha \hspace{2pt} \mathrm{e} ^{-2t/\tau_1} + (1-\alpha) \hspace{2pt} \mathrm{e} ^{-2t/\tau_2}
\end{equation}
Thus for long times one would expect to find:
\begin{equation}
\label{Qlong}
Q(t) \sim (1-\alpha)\hspace{2pt} \mathrm{e} ^{-2t/\tau_2} \sim \frac{C(t)^2}{(1-\alpha)}
\end{equation}
and for short times:
\begin{equation}
\label{Qshort}
Q(t) \sim \alpha \hspace{2pt} \mathrm{e} ^{-2t/\tau_1} + (1-\alpha) \sim
\frac{(C(t)+\alpha-1)^2}{\alpha} + (1-\alpha)
\end{equation}

\begin{figure}[t!]
\subfigure[$\beta=5$, $\alpha=0.0899$ \label{5hqeq}]{\resizebox{!}{170pt}{\includegraphics{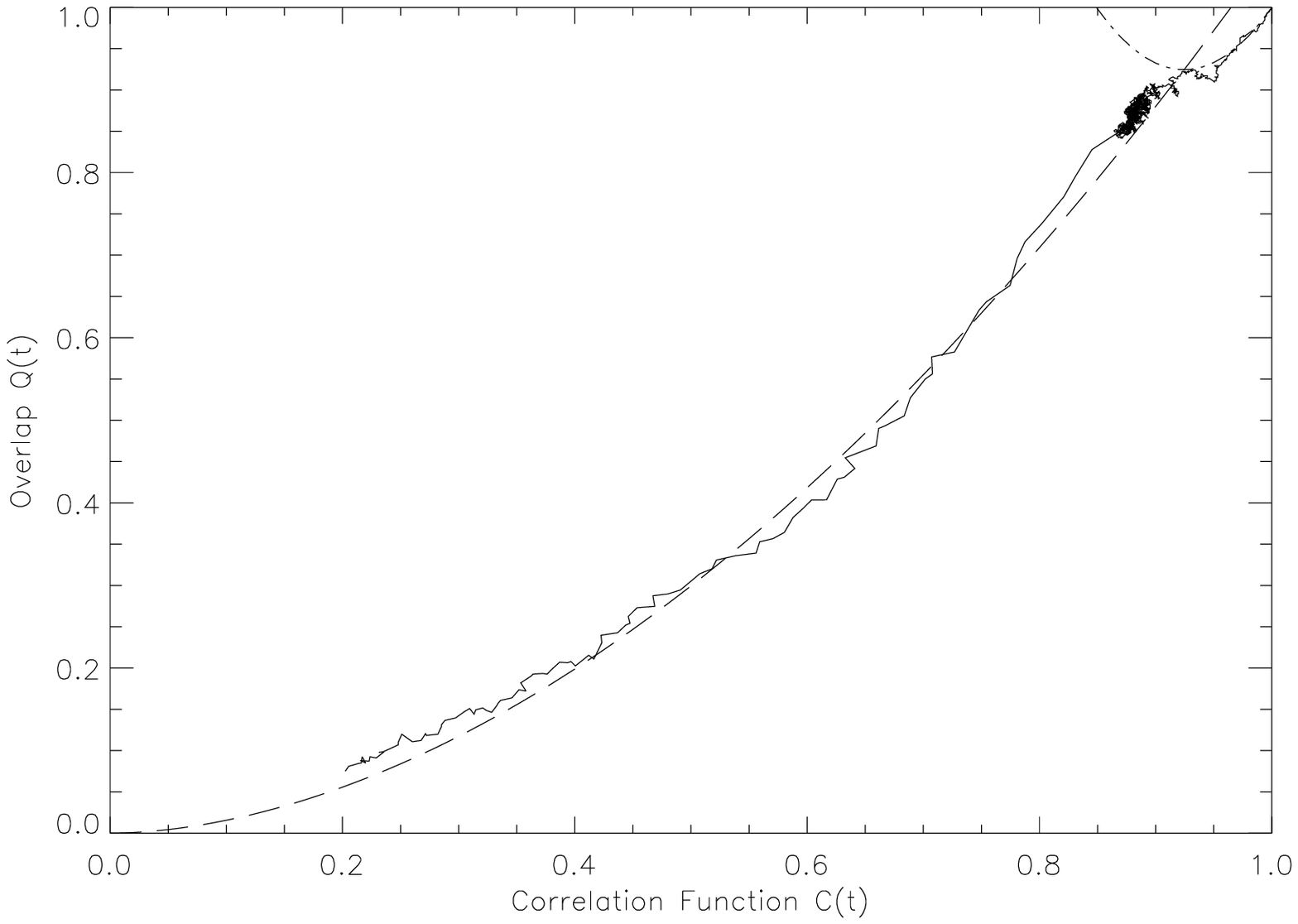}}}
\subfigure[$\beta=4$, $\alpha=0.222$ \label{4hqeq}]{\resizebox{!}{170pt}{\includegraphics{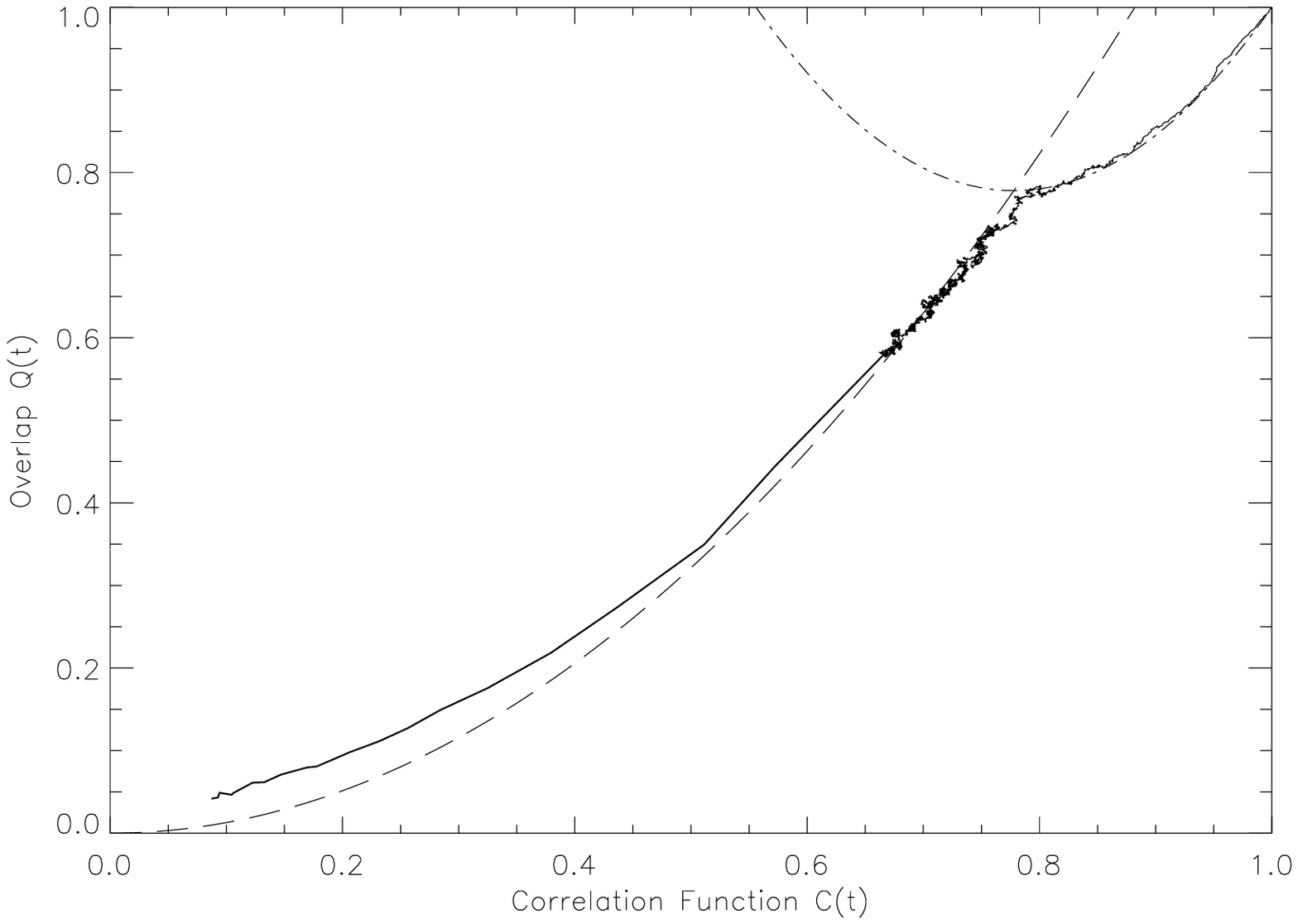}}}
\begin{center}
\subfigure[$\beta=3$, $\alpha=0.447$ \label{3hqeq}]{\resizebox{!}{170pt}{\includegraphics{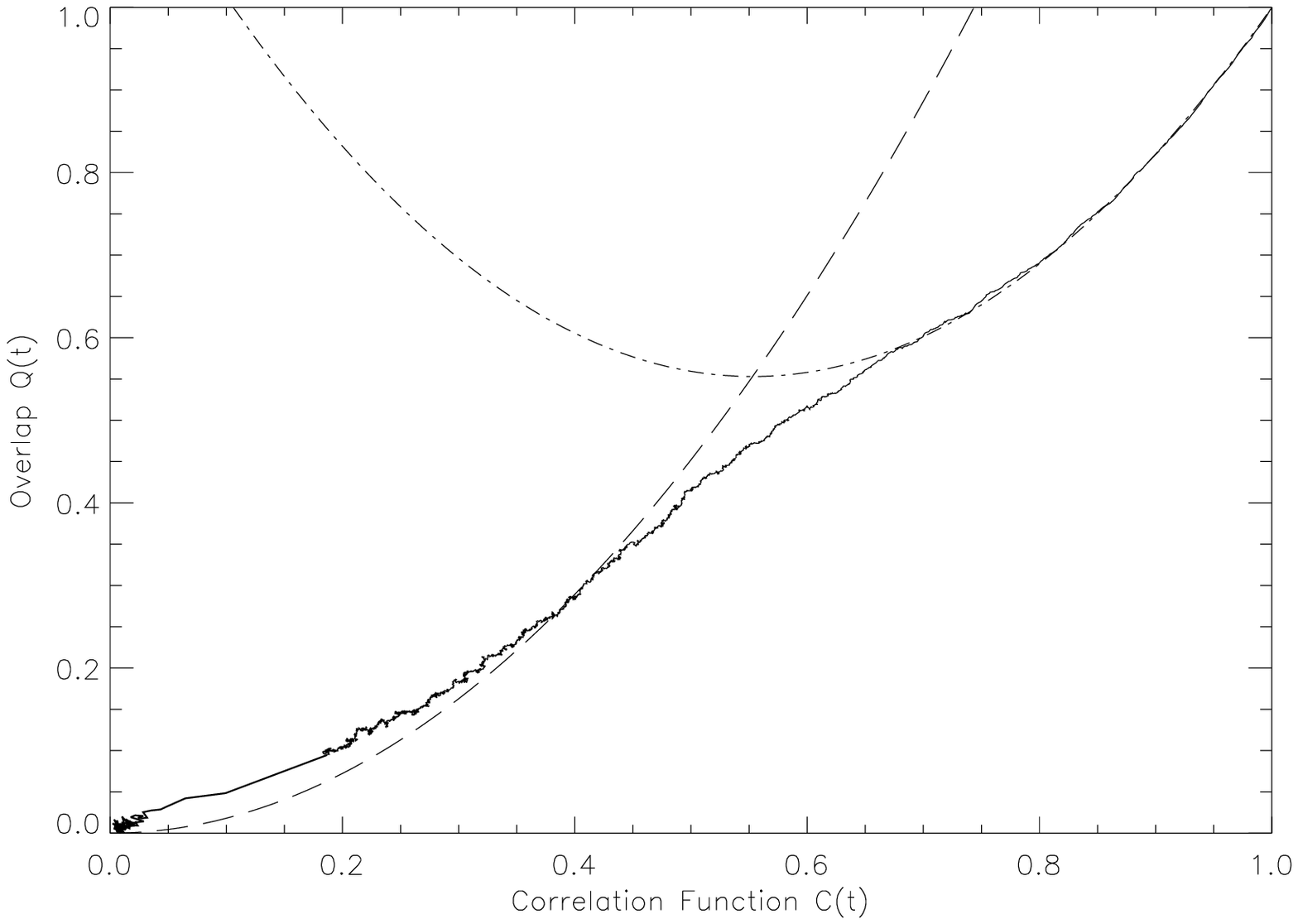}}}
\caption{\textbf{The overlap function $\boldsymbol{Q(t)}$ against the
correlation function $\boldsymbol{C(t)}$.} In each case the dot-dashed curve is the expected short time
behaviour and the dashed curve is the expected long time behaviour, if
equation (\ref{correlfit}) holds. The values of $\alpha$ are
those fitted in the previous section.\label{hqeq}}
\end{center}
\end{figure}
\begin{figure}[t!]
\begin{center}
\resizebox{!}{260pt}{\includegraphics{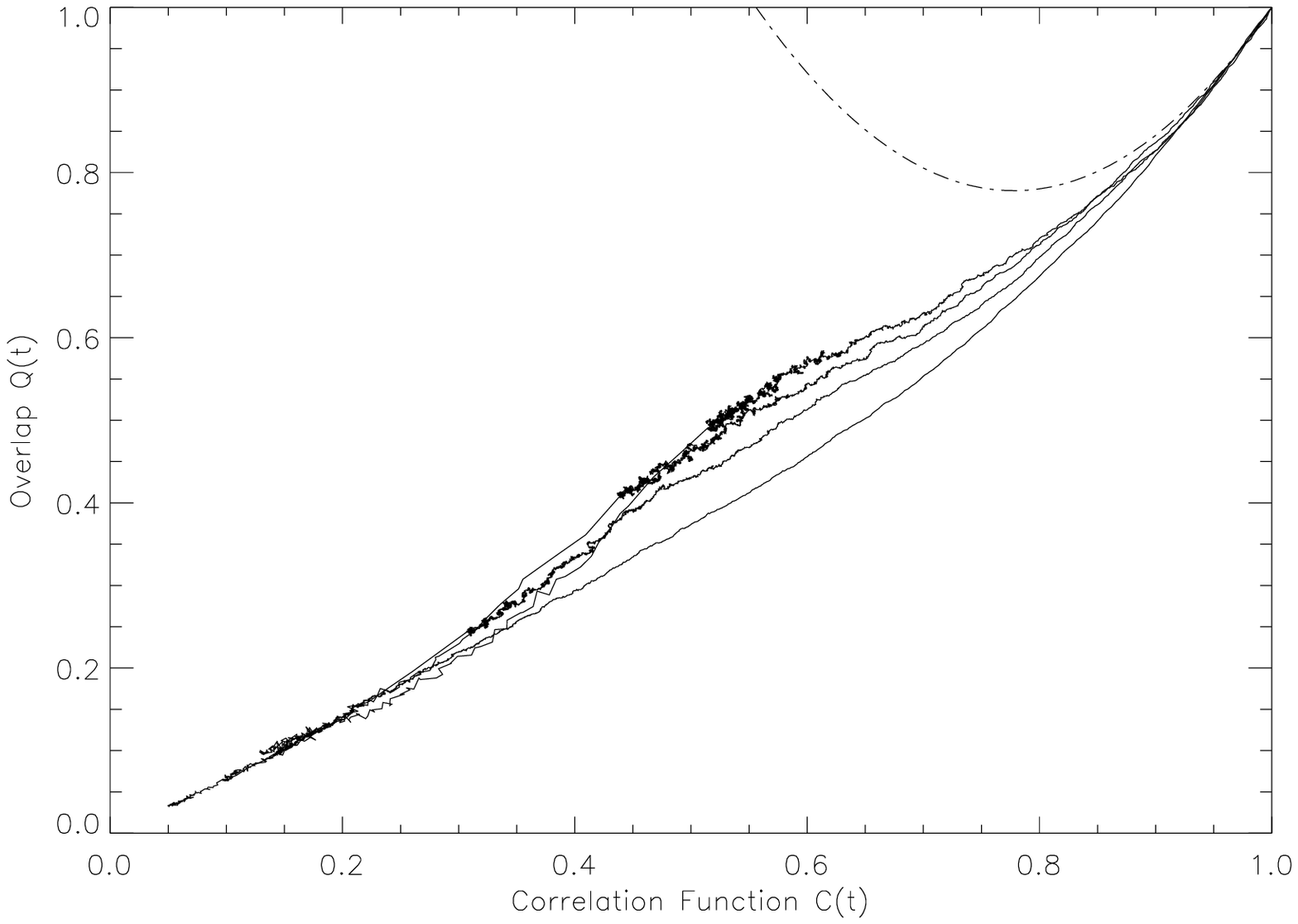}}
\caption{\textbf{$\boldsymbol{Q_{t_w}(t)}$ against
$\boldsymbol{C(t_w, t_w+t)}$ for $\boldsymbol{\beta=4}$.} From
lower to upper curve, $t_w=10N, 10^2N, 10^3N,
10^4N$. The dotted curve is that of expected short time behaviour for
$Q(t)$ against $C(t)$ in equilibrium.\label{4qlap}}
\end{center}
\end{figure}

Figure \ref{hqeq} shows parametric plots of the overlap $Q(t)$ against the
correlation function $C(t)$ (in equilibrium) for different values of
$\beta$. These plots should be `read' from the top right corner
i.e. $t=0$ occurs when $Q(t)=C(t)=1$, and long times correspond to low
values of $Q(t),C(t)$. We have plotted on each the expected short time
and long time behaviour as given in equations (\ref{Qlong}) and
(\ref{Qshort}), 
where the values of $\alpha$ are those fitted in the
previous section. One sees that the short time expression fits the
data extremely well, supporting our hypothesis that the initial decay
of the correlator, from 1 down to the plateau, is exponential. The
longer time behaviour initially fits very well for both $\beta=4$ and
5 (Figures \ref{5hqeq}, \ref{4hqeq}), although as time goes on the theoretical curve drops below the
data. This tends to suggest that at long time-scales there is some
correction to this fitted form which we have not taken into
account; the second relaxation may be some kind of modified
exponential rather than pure exponential. For the case of $\beta=3$ (Figure \ref{3hqeq}) the values of
$\tau_1, \tau_2$ are much closer together and thus the two time-scales
are not so well separated. Therefore one does not see a well-defined
plateau in the equilibrium correlation function (see Figure
\ref{heqcor}), and the cusp in this parametric plot is also not as
clear. The cusp in these plots is a direct result of the existence of
the plateau; since $Q(t)=C(2t)$ in equilibrium, the overlap function
$Q(t)$ reaches the plateau before the correlator $C(t)$. Thus there is
a time period for which $Q(t)$ is effectively stationary whilst $C(t)$ is
still dropping fast. This is followed by a time period for which the
plateaux in both functions overlap, and therefore both are stationary,
and then there is a regime  in which $Q(t)$ drops away from the plateau
whilst $C(t)$ is still stationary. This tells us that whilst the
dominant process is diffusion of the dimers, the two copies of the
system are restricted to a narrow area of phase space; this is because
the isolated defects have not yet moved in either copy, and thus the
overlap will be high. It is only once the activated processes become
dominant that the two copies can move well apart from each other.

Figure \ref{4qlap} shows a parametric plot of the overlap against the
correlator for the non-equilibrium case i.e. $Q_{t_w}(t)$
against $C(t_w,t_w + t)$. The initial behaviour is independent of
$t_w$ and in fact follows the short time behaviour we expect in
equilibrium. However, as $C(t_w,t_w + t)$ drops below 0.9, the
curves fall below the equilibrium behaviour and we see evidence of dependence on $t_w$. The cusp that develops is sharper for larger
$t_w$, with the behaviour tending towards that of equilibrium. For
the smaller values of $t_w$ there is no channelling in phase space
because the starting configuration is well away from equilibrium, and
thus there are many different energetically favourable routes to be
taken. It is clear from these figures that our model falls into the
class of Type II, as the overlap decays to zero rather than a finite
value as it would do for Type I systems; this is as expected as we
believe our model to be glassy rather than dominated by coarsening.

To summarise our findings so far, we have shown that the $D>0$ model
does indeed give
results that are quantitatively similar to those of the topological
model, whilst having the advantage of being simpler, computationally
faster and more suited to analytic study. We have developed a
conceptual picture involving both fast and slow dynamics, and many of
the features of this model can be described in terms of this
picture. We find good agreement between theoretical predictions and
data for the behaviour of the correlation function, energy and overlap function.

\section{$\boldsymbol{D<0}$}
\subsection{Relaxation dynamics and correlation functions}
\begin{figure}\phantom{*}
\begin{center}
\subfigure[The behaviour of the energy under cooling ($D<0$). The values of $t_w$ are the waiting times at each point.]{\resizebox{!}{265pt}{\includegraphics{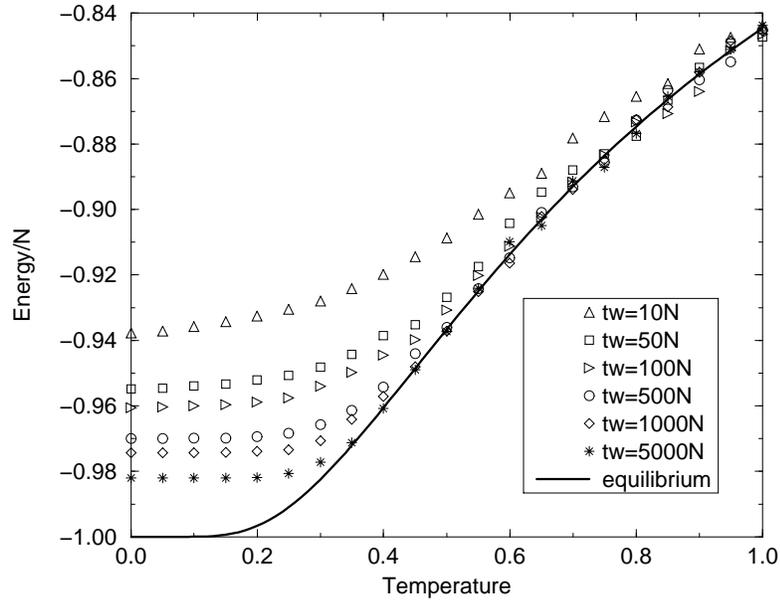}}\label{pcool}}
\subfigure[The behaviour of the energy after a rapid quench
($D<0$). The values of $t_w$ are the times, subsequent to the quench, at
which the energy is measured.]{\resizebox{!}{265pt}{\includegraphics{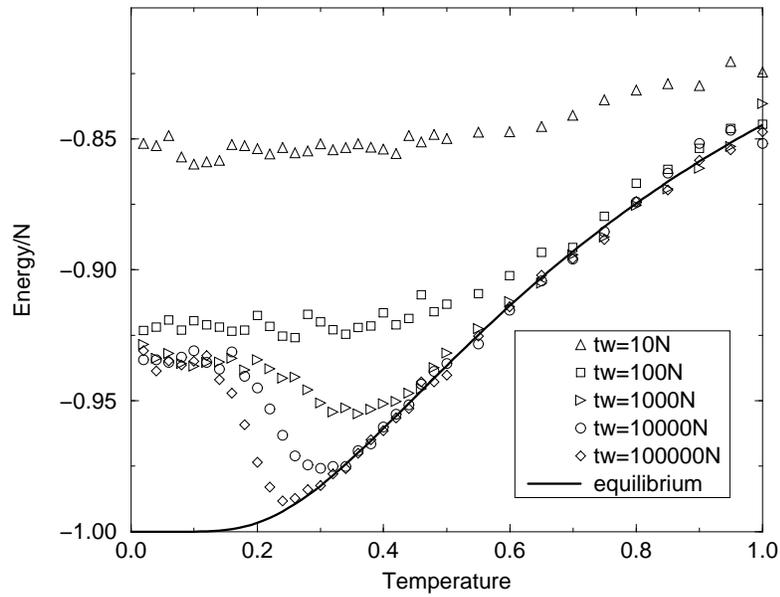}}\label{pquench}}
\caption{\textbf{Energy against temperature for slow cooling and rapid
quench.} \label{png}}
\end{center}
\end{figure}
\begin{figure}\phantom{*}
\begin{center}
\subfigure[Energy against time.]{\resizebox{!}{268pt}{\includegraphics{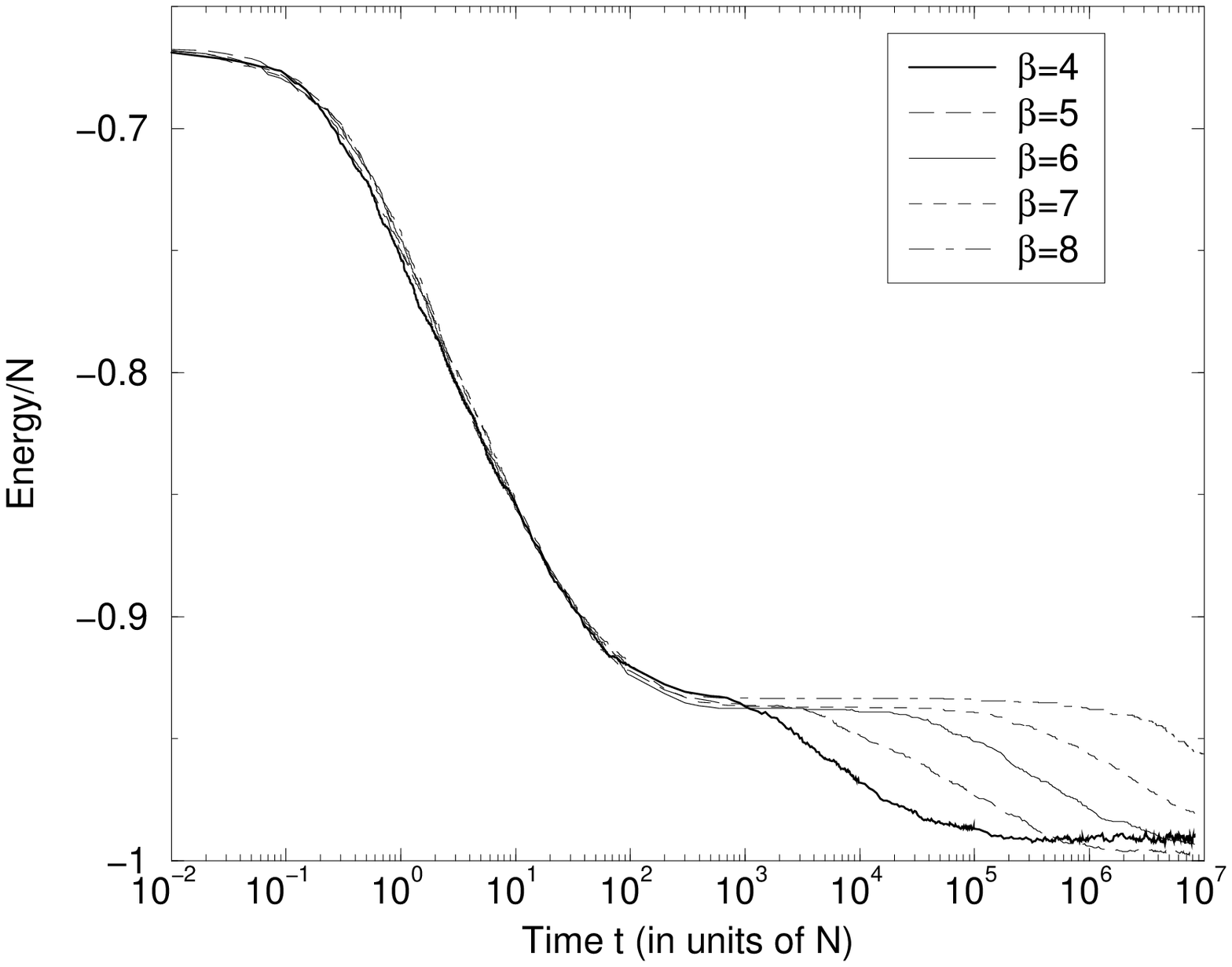}}\label{pnrj}}
\subfigure[Energy against $T \hspace{1pt}\ln t$, where $t$ is measured in units of $N$.]{\resizebox{!}{268pt}{\includegraphics{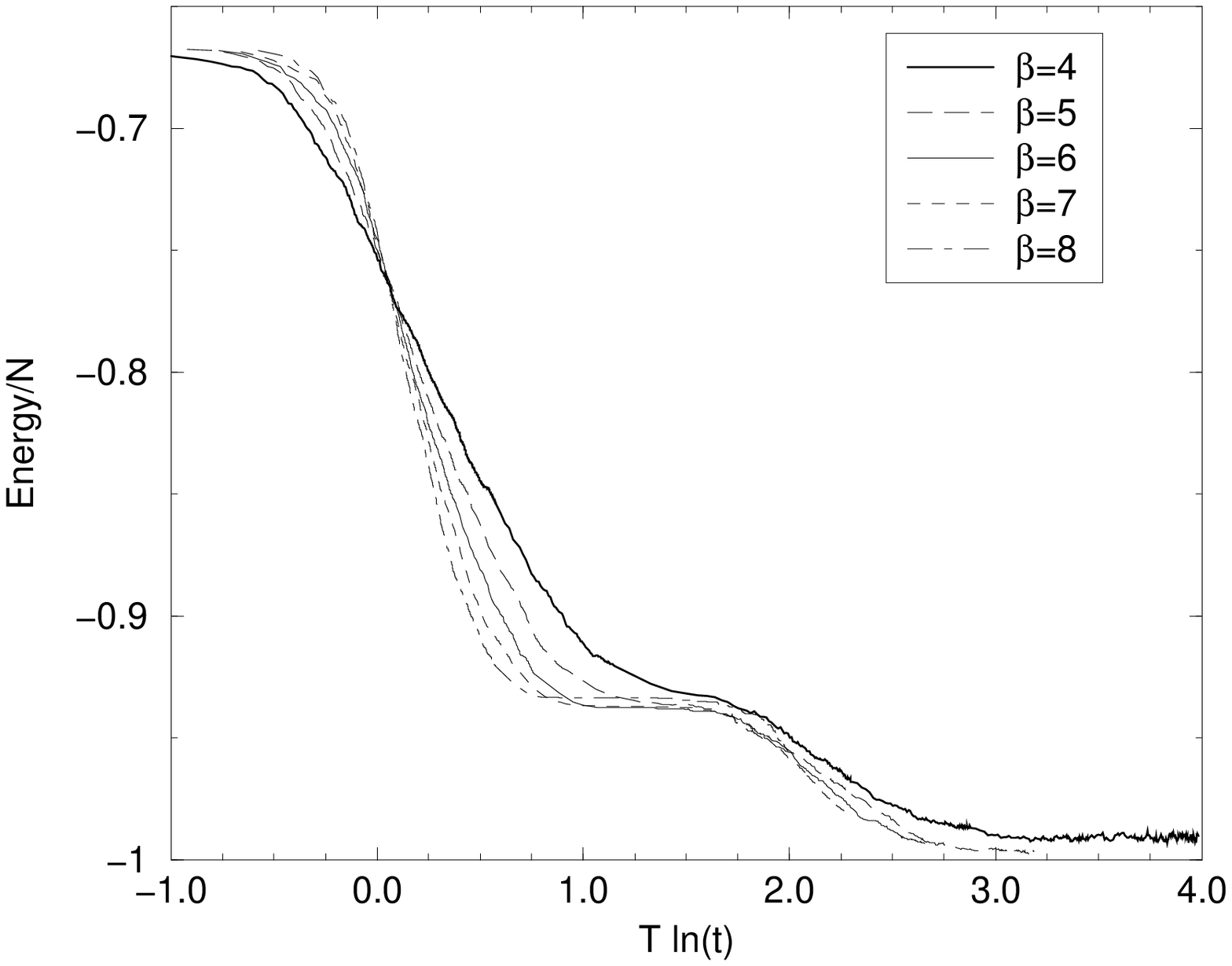}}\label{plognrj}}
\caption{\textbf{The behaviour of the energy with time.} \label{pengy}}
\end{center}
\end{figure}
\begin{figure}[t]
\subfigure[The persistence of non-isolated zero-spins for $D<~0$.]{\resizebox{!}{180pt}{\includegraphics{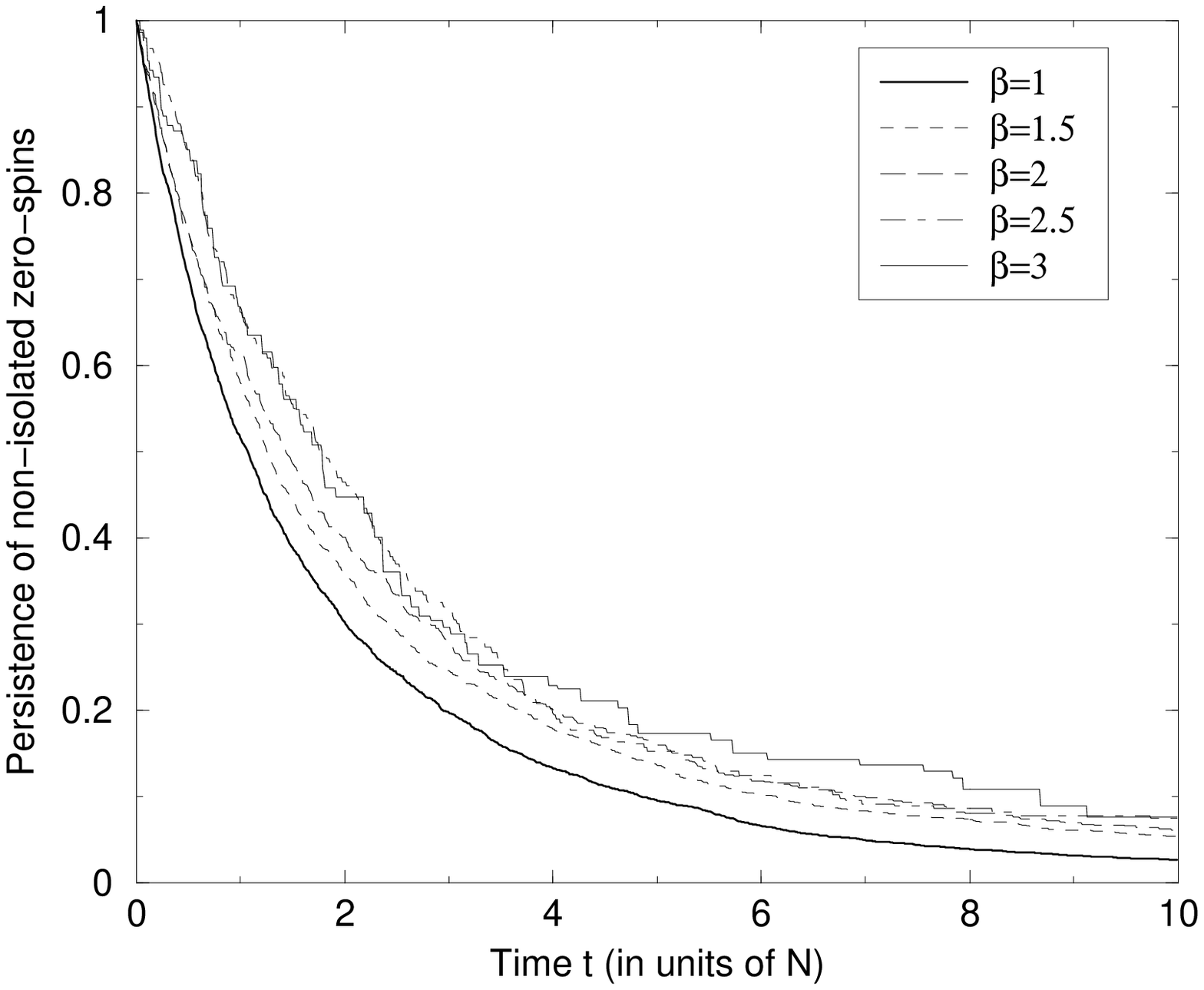}}\label{pperst}}
\subfigure[The persistence of non-zero spins with at least one
opposite neighbour for $D>~0$.]{\resizebox{!}{180pt}{\includegraphics{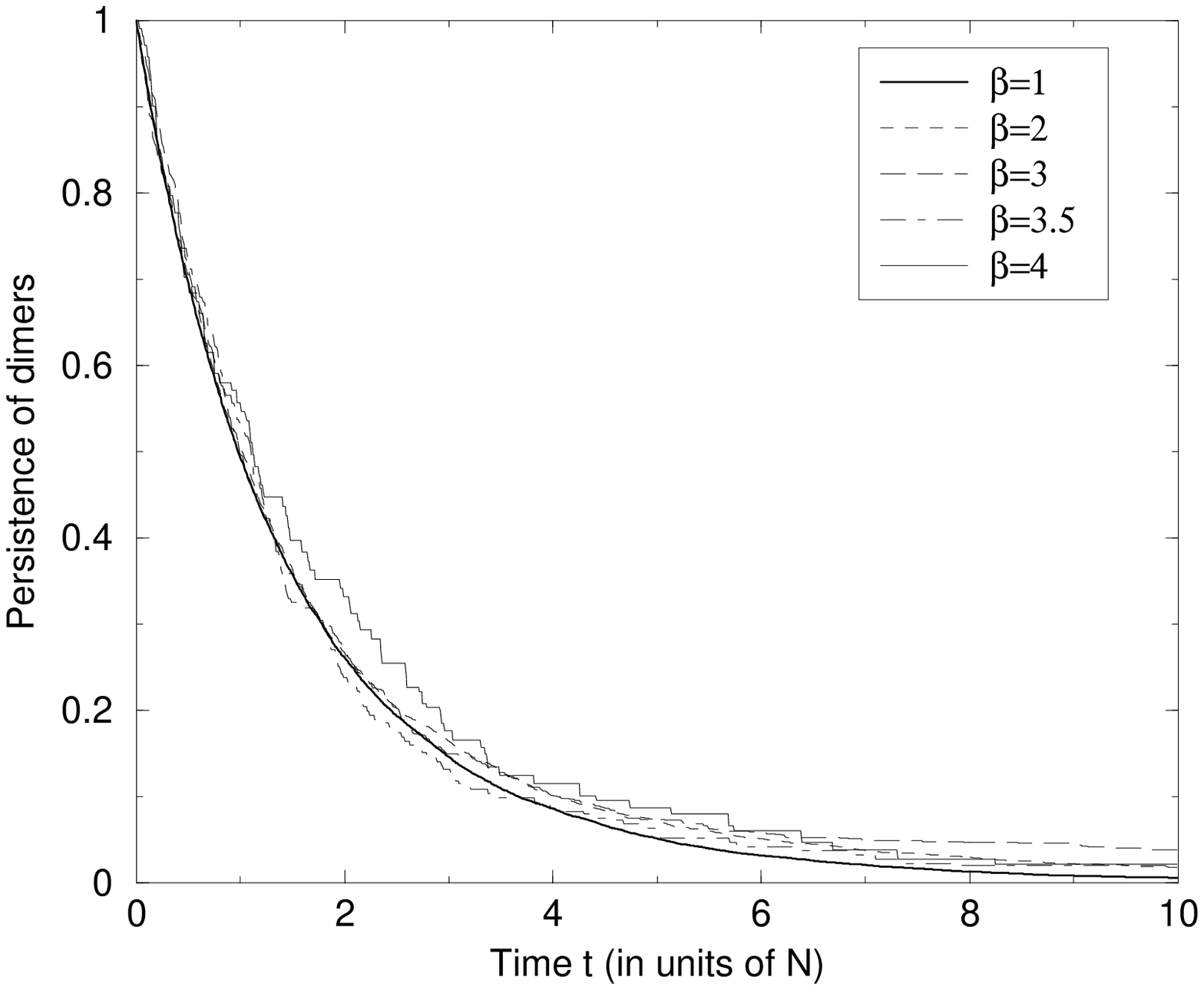}}\label{hperst}}
\caption{\textbf{Persistence functions for both $\boldsymbol{D<0}$ and
$\boldsymbol{D>0}$.}\label{persist}}
\end{figure}

We shall now turn our attention to the model with $D<0$, setting $D=-1$. Unlike the
previous case, this model does not have a unique absorbing ground
state; instead, there are a great many degenerate ground states,
although some of these are not accessible using our dynamical
rules. This raises the possibility that in preparing an `equilibrium'
system by randomly placing throughout the lattice the correct number of $\pm 1$'s for that
temperature, one might pick out an inaccessible
configuration. However, the probability of this occurring is so small
as to be negligible, and we have checked that the results obtained
by this method do not differ from those obtained through waiting long
enough for equilibration to occur. Thus when we refer to an
equilibrated system, we mean one that has been prepared through a
random allocation of the correct number of non-zero spins.

We expect the $D<0$ model to also show glassy behaviour with two
different time-scales, but as the system now
favours $s_i=\pm 1$ rather than $s_i=0$, the equivalent of the
free-moving dimers will be fast-moving pairs of spin zero, and the
analogue of the slow-moving energetically-trapped defects will be isolated zero-spins. With the correct choice of
observables, one expects to see all the same features as with the
$D>0$ model. However, although one expects similar qualitative
behaviour,\footnote{except in pathological cases} the quantitative behaviour should be different - this is
because $A + A \to \emptyset$ processes replace those of type $A + B \to \emptyset$ and also
because the zero-spin dimers cannot move quite so easily through the
$\pm 1$ background. As shown in Figure \ref{stuck} there are certain
configurations that simply cannot move. 

Simulations were again performed for $N=9900$. Figure \ref{pcool} confirms that we do see glassy behaviour when the
system is cooled at different cooling rates. A plot of the energies attained
after running for a variety of times $t_w$ at various temperatures $T$ from a
starting configuration corresponding to infinite temperature are
shown in Figure \ref{pquench}; the presence of activated processes is
indicated by the clear minima and by the plateau at $\frac{E}{N} \sim
-0.94$ below which the system cannot penetrate at low temperatures,
even after the longest waiting times. Again one can see this plateau clearly
in a plot of energy against time subsequent to a quench from infinite
temperature (see Figure \ref{pnrj}); one can again re-scale the time
axis to $T \ln t$ in order to see the staircase shape appear (Figure
\ref{plognrj}).

Because the background does not necessarily allow the dimers to move
freely, one must consider the possibility that the dimer diffusion may
be dependent upon the density of non-zero spins. In equilibrium, this
density is dependent upon the temperature; therefore we have investigated the
persistence of non-isolated zero-spins under
equilibrium conditions. In order to do
this, one can identify all zero-spins that have at least one
neighbouring spin which is also zero in the starting equilibrium
configuration; as the system evolves, one can measure the fraction of
these that have NOT been involved in a move. Figure \ref{pperst} shows these
results for $D<0$; one can see that the persistence is weakly temperature-dependent. One cannot sensibly investigate the effect at lower
temperatures because there are so few zero-spins present in equilibrium conditions. For comparison, we also show the results for the $D>0$
model, where the persistence is defined as the fraction of non-zero
spins with at least one neighbour of the opposite sign that have NOT
undergone a move. The results are shown in Figure \ref{hperst} and
show no temperature-dependence. This result has implications for the
form of both the correlation function and the energy.
\begin{figure}
\begin{center}
\subfigure[The energy for (from left to right) $\beta=5,6,7$ fitted
with equation (\ref{nrjcrap}); the fits are the dashed curves. The parameters
($a_1,m_1,\kappa_1,\kappa_2$) are as as follows: for $\beta=5$,
(-0.942,1.02,0.49,0.52); for $\beta=6$,(-0.941,0.93,0.50,0.51) and for
$\beta=7$, (-0.941,0.93,0.49,0.54).]{\resizebox{!}{245pt}{\includegraphics{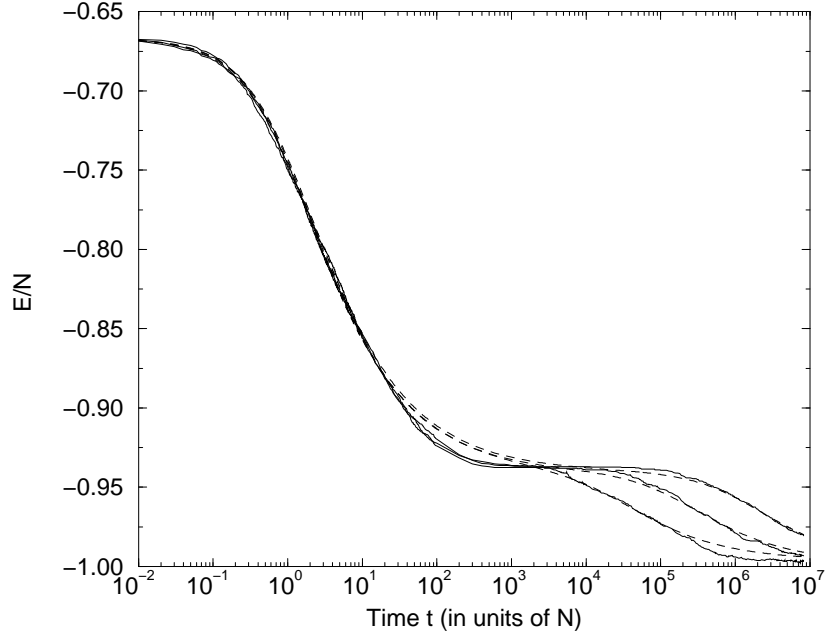}}\label{pbadfit}}
\subfigure[The energy fitted with equation
(\ref{negefit}); the fits are the solid
lines.~{($a_1,a_2,m_1,\kappa_1,m_2,\nu,\kappa_2$) are
as~follows~:~for~$\beta$=5,}
(-0.936$,$0.61,0.748$,$0.98,0.0385$,$0.638,0.62);~for~$\beta$=6,
(-0.937,0.57$,$0.77,0.977$,$0.0389,0.64,0.59) and for~$\beta$=7, ($-0.937,0.66,0.46,1.0,0.0313$,$0.474,0.59$).]
{\resizebox{!}{245pt}{\includegraphics{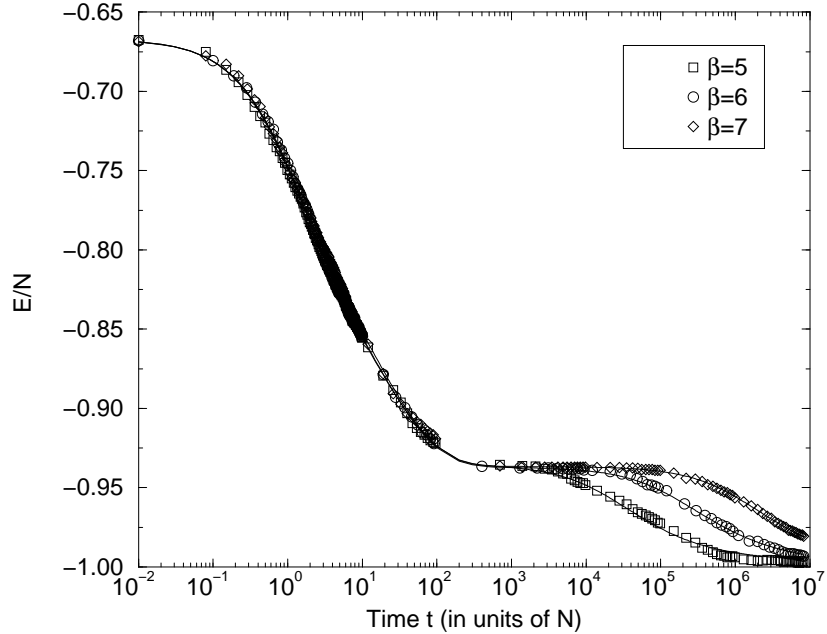}}\label{pefit}}
\caption{\textbf{Fits to the energy.}\label{pfits}}
\end{center}
\end{figure}

In order to produce a fit for the energy relaxation from a starting
configuration corresponding to infinite temperature, one must think very clearly
about the processes involved. The initial fast decay to the plateau
involves two pairs of zero-spin dimers annihilating; this is of type
$A + A \to \emptyset$, which one expects to give a $t^{-\frac{d}{2}}$ dependence
in the energy in the asymptotic limit \cite{hinrich, mattis}
\footnote{In fact the critical dimension $d_c=2$ for this theory, and
one expects logarithmic corrections at this point, but we shall ignore these.}.  At this stage the density-dependence of the dimer
diffusion is not likely to be strong enough to cause substantial
deviation from the $A + A \to \emptyset$ behaviour so one expects the initial decay to
behave as  $(1 + m_1t)^{-1}$; however, it is not trivial in this case to determine
what the value of $m_1$ might be. Because the zero-spin dimers are not
completely free to move diffusively through the background  we are
unable to produce an accurate estimate; we simply expect a value of
$m_1$ of the order of 1. For the slow decay from the
plateau to equilibrium, the density-dependence of the dimer diffusion
will have a substantial effect  because there are far more $\pm 1$
spins present as the system flows closer to equilibrium. The slow
process involves the pairing of isolated zero-spins through the
mechanism of dimer creation and absorption; it is unlikely to behave
exactly as an $A + A \to \emptyset$ process given that the dimers cannot
diffuse freely to facilitate it. One can fit this latter part
separately, and one finds it takes the form
$a_1 (1+t\mathrm{e}^{-2\beta})^{-\kappa}$; the factor of two in the
exponential was fitted as a free parameter and a result of almost
exactly 2 was obtained for every temperature. This is in keeping with the
energy barrier of $2$ involved in creating a dimer. The parameter
$a_1$ is naturally associated with the plateau value and clearly takes
a value $\sim -0.94$. The value of
$\kappa$ is approximately $0.6$; this is substantially slower than the behaviour
one would find asymptotically with a pure $A + A \to \emptyset$ process and is
due to the inhibited movement of the dimers.

Having fitted the latter part, we attempted a fit of the full dataset
of the form:
\begin{equation}
\label{nrjcrap}
\frac{E}{N}=\left(-\frac{2}{3} - a_1 \right)
\left(1+m_1 t\right)^{-\kappa_1} +
\left(a_1-e_{eq} \right)\left(1+t\mathrm{e}^{-2\beta} \right)^{-\kappa_2} + e_{eq}
\end{equation}
where $a_1,\kappa_1,\kappa_2, m_1$ are all parameters to be
determined, and $e_{eq}$ is the equilibrium energy per cell at the temperature
in question. One cannot fit this well to the data, as shown in Figure
\ref{pbadfit}: the decay is faster than that of a power law in the latter
stages of the decay to the intermediate plateau. Thus we must think
again about the processes involved in the relaxation of the energy to
the plateau.

As mentioned earlier, besides the $A + A \to \emptyset$ fast processes, there
is also a fast process involving a dimer interacting with a defect to
leave an isolated defect - this is of type $A + C \to \emptyset + C$, and thus
typically gives a stretched exponential for the asymptotic behaviour
of the energy with time. We did not need to include these in the fit
to the energy for $D>0$; however, it is clear from the poor fit in
Figure \ref{pbadfit} that in the $D<0$ case we cannot neglect them. We
therefore expect the energy density $E/N$ to be approximated by the following
form:
\begin{equation}
\label{negefit}
\frac{E}{N} = \left(-\frac{2}{3} - a_1\right)\left( a_2
\left(1+m_1t\right)^{-\kappa_1} + (1-a_2) \mathrm{e}^{-\left(m_2 t \right)^{\nu}}\right) +
\left (a_1-e_{eq} \right)\left(1+t\mathrm{e}^{-2\beta} \right)^{-\kappa_2} + e_{eq}
\end{equation}
where $a_1,\kappa_2$ can be fitted separately from the decay from the
plateau to equilibrium, and $a_2, \kappa_1,$ $\nu, m_1, m_2$ are
parameters to be determined. The theory would suggest that $\kappa_1$
should be close to 1, and that $m_1$ should be of the order of
1. Figure \ref{pefit} shows these fits superimposed on the data; it
will be noted that agreement is excellent. The values of the
parameters are given in the caption; in particular, one notes that
$\kappa_1$ is extremely close to 1 in each case, and that $m_1$ is
indeed of the order of 1. We have already discussed the power law
decay from the intermediate plateau to equilibrium. 
Without developing a much more complex
theory one can say little about the parameters in the stretched
exponential term.
\begin{figure}[t]
\begin{center}
\resizebox{!}{270pt}{\includegraphics{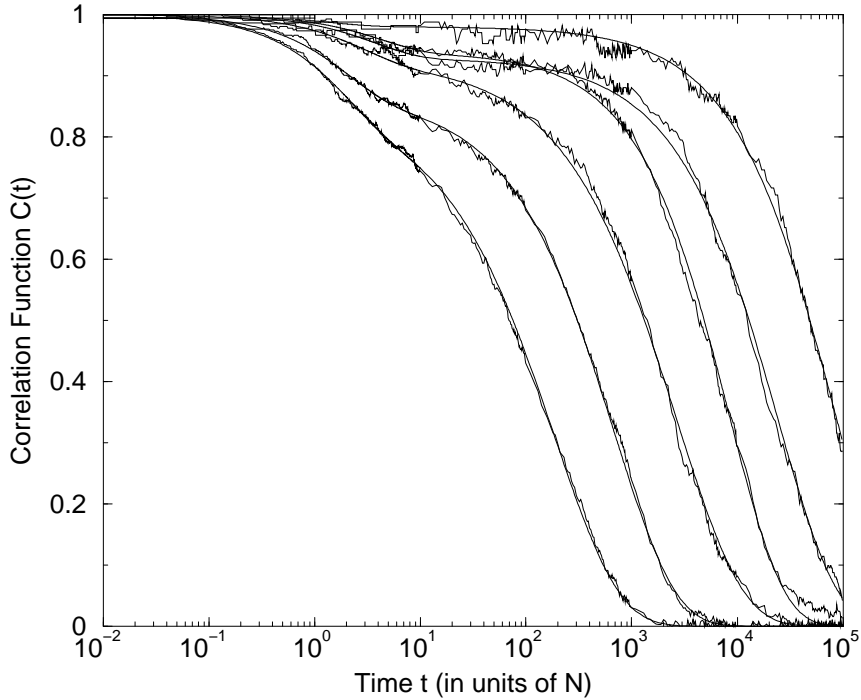}}
\caption{\textbf{Correlation functions in equilibrium conditions.} From
left to right, $\beta=2.5,3,3.5,4,4.5,5$. The superimposed fits are of
the form: $C(t) = \alpha \hspace{2pt}\mathrm{e}^{-t/\tau_1} + (1-\alpha)\hspace{2pt}\mathrm{e}^{-(t/\tau_2)^\gamma}$. \label{pcor}}
\end{center}
\end{figure}

With regards to the correlation function, in order to study the $D>0$
model we chose a function which focussed on the defects. We do
the same in this case, and thus the correlation function we measure is:
\begin{equation}
\label{pcorrel}
C(t_w, t_w+t) =\frac{1}{1-p_{eq}(0)} \left(\frac{ \sum_{i=1}^N \delta_{s(t_w),0} \hspace{1pt}
\delta_{s(t_w + t), 0}}{\sum_{i=1}^N \delta_{s(t_w),0}}-p_{eq}(0)\right)
\end{equation}
where $p_{eq}(0)$ is the equilibrium density of zero spins at the
temperature of interest. The somewhat unusual normalisation is
necessary to allow the correlation function to decay to zero rather
than to a finite value equal to $p_{eq}(0)$; this finite plateau comes about
because a fraction $p_{eq}(0)$ of the cells that have spin zero at time $t=0$
can be expected to have spin zero at any later time. In the $D>0$
model, we did not have to take into account this effect since the
contribution from spins that remain $+1$ (or -1), and those that swap
from +1 to -1 (or -1 to +1) cancel out. Figure \ref{pcor} shows that
this function does indeed produce similar results to those of the
$D>0$ model: again, we can see clear evidence of two-step
relaxation. The fits superimposed upon the data are of the following
form:
\begin{equation}
\label{stretch}
C(t) = \alpha \hspace{2pt}\mathrm{e}^{-t/\tau_1} + (1-\alpha)\hspace{2pt}\mathrm{e}^{-(t/\tau_2)^\gamma} 
\end{equation}
\begin{figure}[t!]
\subfigure[$\tau_1$ against inverse temperature]{\resizebox{!}{180pt}{\includegraphics{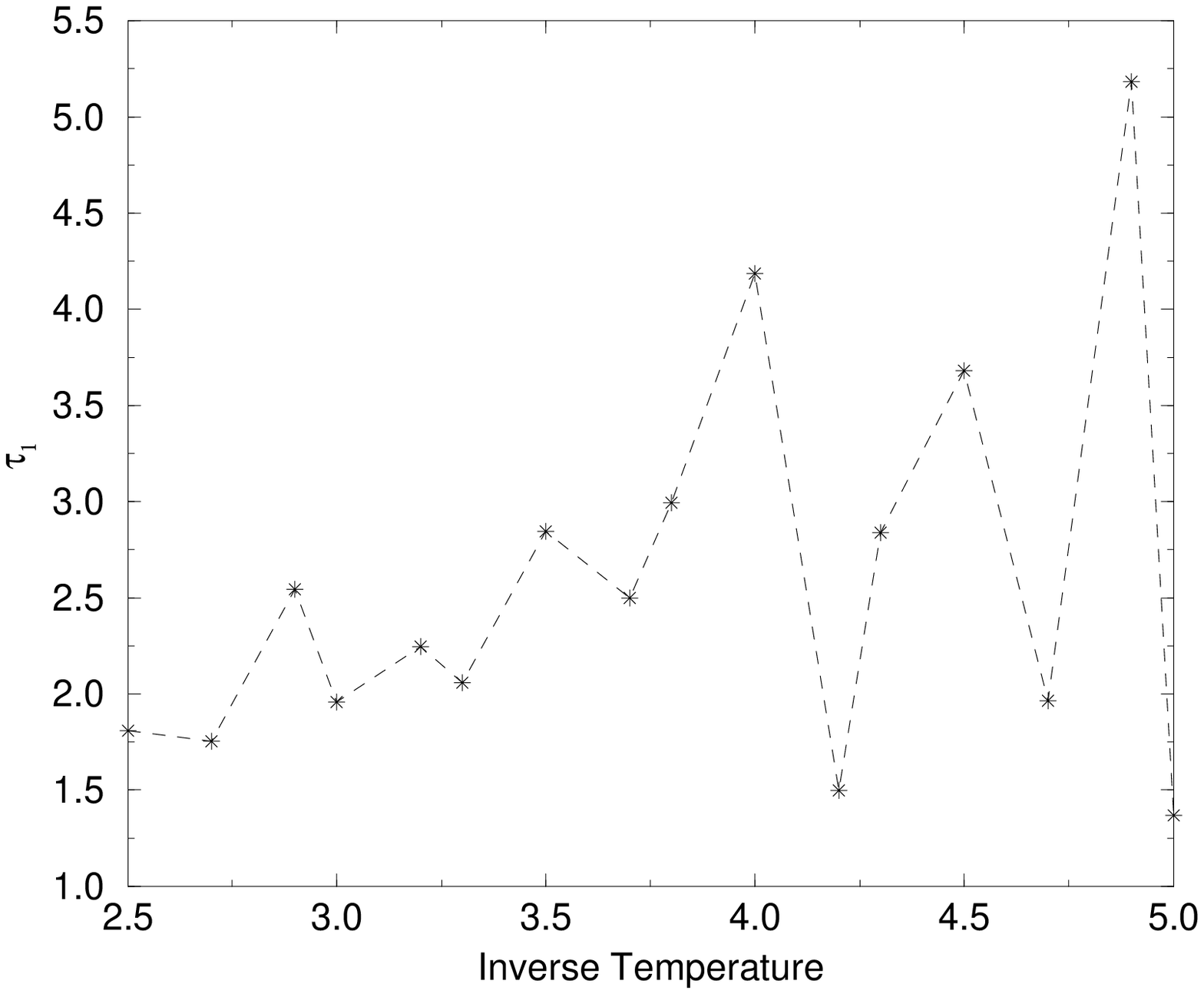}}\label{pst1}}
\subfigure[$\tau_2$ against inverse temperature]{\resizebox{!}{180pt}{\includegraphics{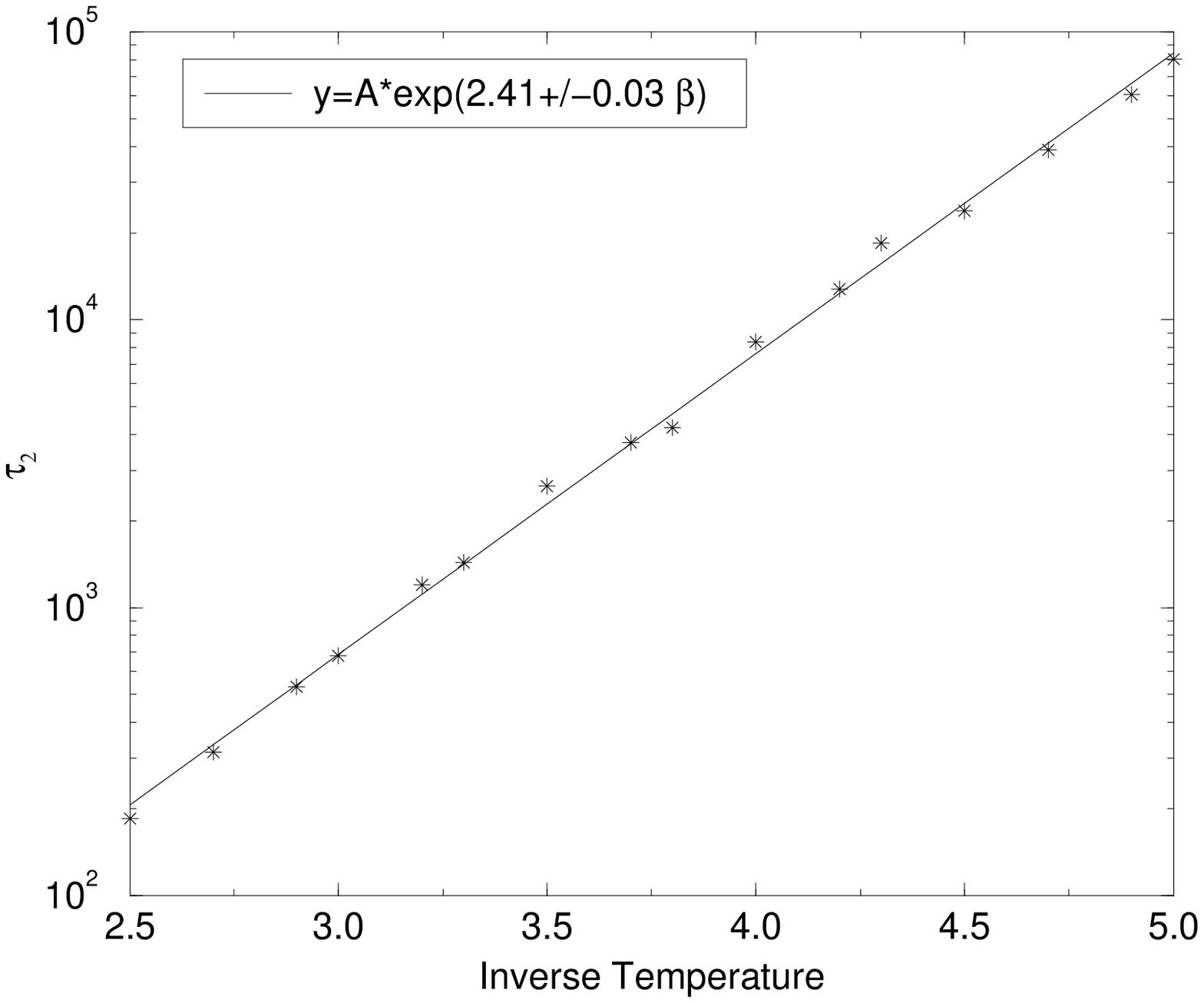}}\label{pst2}}
\subfigure[$\alpha$ against inverse temperature; the superimposed line is $y=~\frac{12\mathrm{e}^{\beta}}{(1+2\mathrm{e}^{\beta})^2} - \frac{\mathrm{e}^{-\beta}}{2}$]{\resizebox{!}{180pt}{\includegraphics{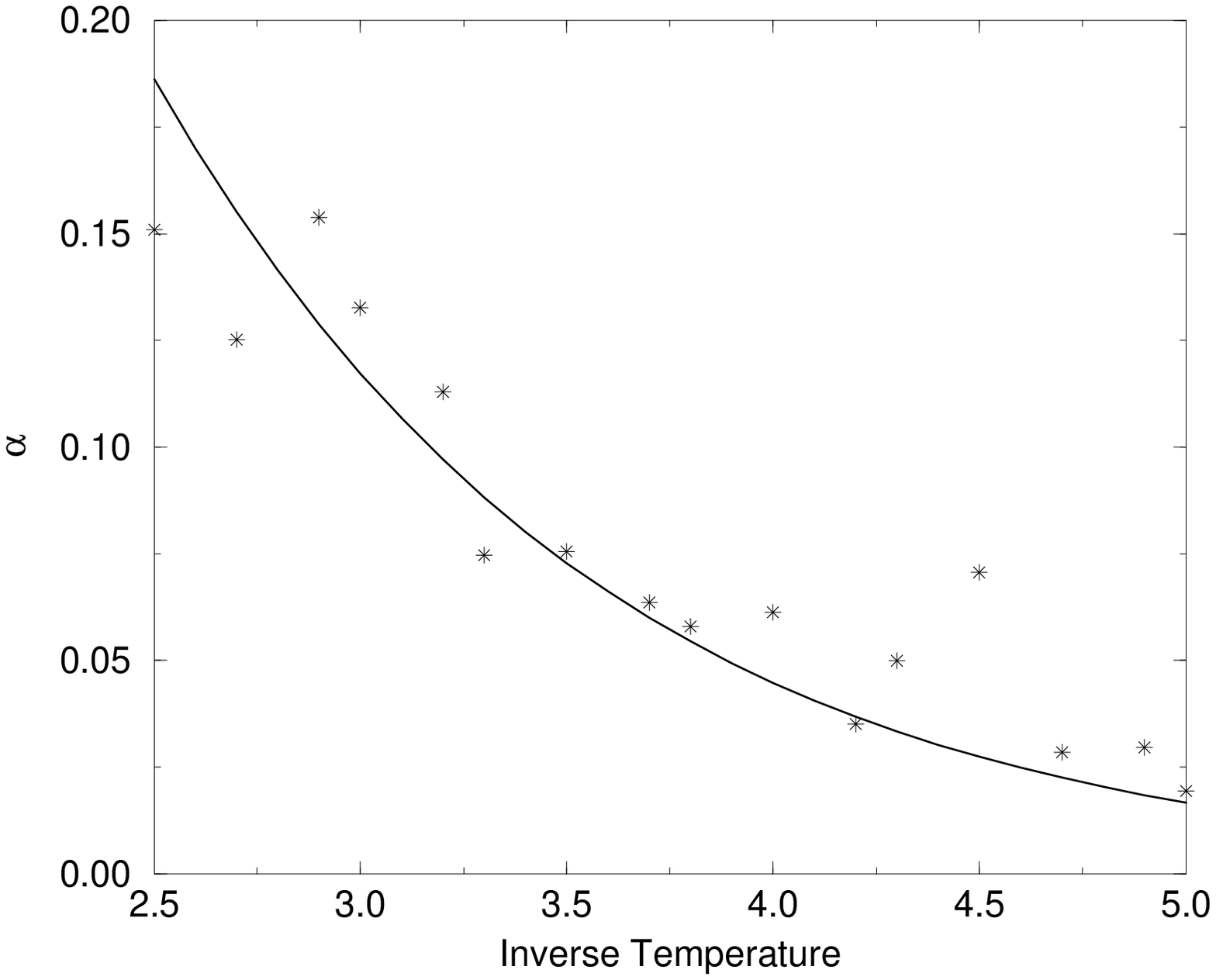}}\label{pplat}}
\subfigure[$\gamma$ against inverse temperature]{\resizebox{!}{180pt}{\includegraphics{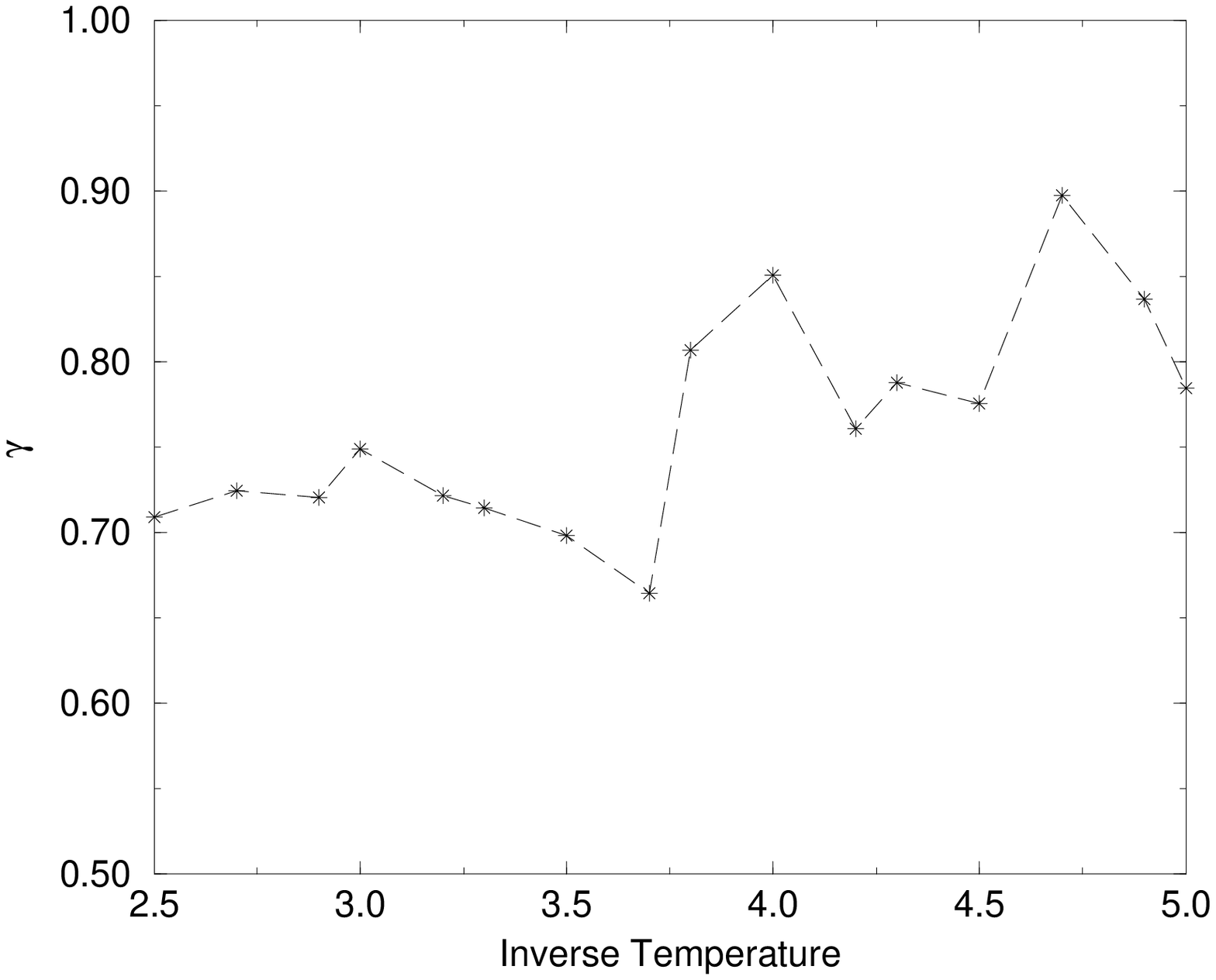}}\label{pexp}}
\caption{\textbf{The behaviour of the fitting parameters $\boldsymbol{\alpha,\tau_1,
\tau_2}$ and $\boldsymbol{\gamma}$ with temperature.} The data was obtained by fitting the
form $C(t) = \alpha \hspace{2pt}\mathrm{e}^{-t/\tau_1} +
(1-\alpha)\hspace{2pt}\mathrm{e}^{-(t/\tau_2)^{\gamma}}$ to the equilibrium
correlation functions.}
\end{figure}
It will be noted that these fits are extremely good, and also that the
relaxation in the long-time region follows a stretched exponential
rather than an exponential as in the $D>0$ case. This is a result of
the temperature dependence of the dimer diffusion, which slows the
decay from the plateau. Thus $\gamma$ in Figure \ref{pexp} is less
than one in all cases. We do not have a full theory of the behaviour of the
system in this region and thus cannot comment further on the
behaviour of this exponent.

One might naively expect the predicted value of $\tau_1$ to be altered
by the fact that the zero-spin dimers are not
necessarily free to move through the $\pm 1$ background. In fact this is not the case: $\tau_1$
is the time-scale for the dimers that are \textit{free to move} only, and we
can still expect those free dimers to move with a time-scale of 2,
independent of temperature, exactly as in the $D>0$ case. Figure \ref{pst1} shows the values of $\tau_1$ against
inverse temperature obtained from fitting the correlation functions
with equation (\ref{stretch}) - this data is in keeping with a
temperature-independent value of $\tau_1 =2$. Those dimers that are
not free to move do not contribute to the initial fast decay of the
correlation function.  The parameter that the extra jamming \textit{does}
alter is $\alpha$: one expects the correlation function to decay to a
plateau value which is one minus the density of free zero-spins
(although one must remember to normalise correctly as in equation (\ref{pcorrel})). Thus
we have to calculate the probability of obtaining a zero-spin dimer
which can move (shown in Figure \ref{diff}, but one should now think
of the $\pm 1$'s as being the background and the zero-spins as being the
dimer), and also of obtaining a zero-spin pair which can
oscillate (as in Figure \ref{oscill}). This gives a probability of
$24p(0)p(1)^2$; thus after normalisation, we expect $\alpha$ to behave as:
\begin{equation}
\label{sta}
\alpha=\frac{12\mathrm{e}^{\beta}}{(1+2\mathrm{e}^{\beta})^2} -
\frac{\mathrm{e}^{-\beta}}{2}
\end{equation}
Figure \ref{pplat} shows this curve superimposed upon the fitted
values of $\alpha$ against temperature; the data shows reasonable agreement
with the theory.

One can see from Figure \ref{pst2} that we again observe Arrhenius
behaviour for $\tau_2$. In this case, however, the best-fit Arrhenius
law is $\tau_2 \sim \mathrm{e}^{2.4\beta}$, whereas the energy barrier
argument would suggest $\tau_2 \sim \mathrm{e}^{2\beta}$. As mentioned
earlier, we lack a full understanding of the behaviour in this
region, and can only say that this discrepancy is probably also due to
the fact that the $\pm 1$ background does not allow the zero-spin
dimers to diffuse freely.

\subsection{Response and overlap functions}
\begin{figure}[t!]
\begin{center}
\subfigure[$\beta=2.5$]{\resizebox{!}{170pt}{\includegraphics{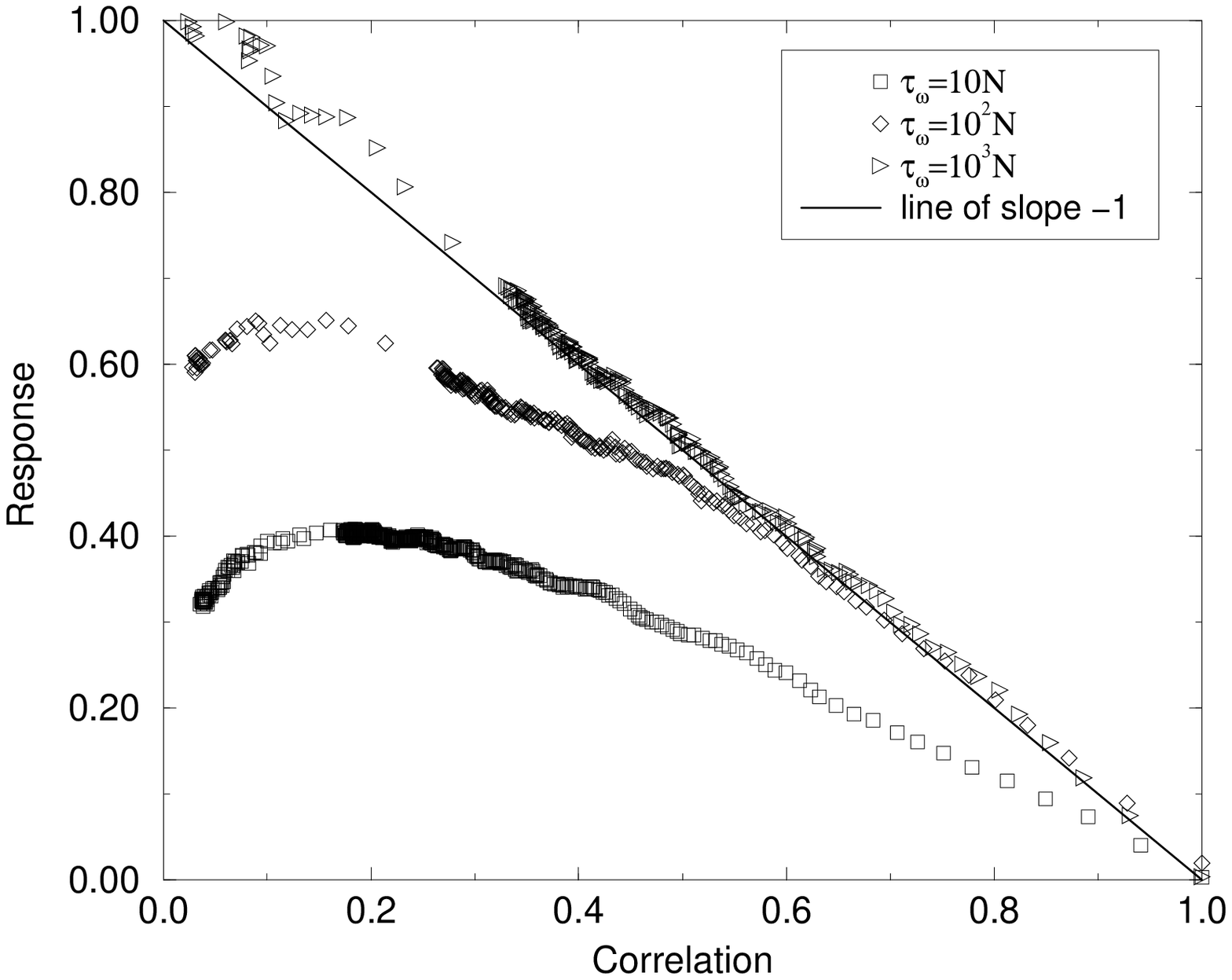}}\label{2.5presp}}
\subfigure[$\beta=4.5$]{\resizebox{!}{170pt}{\includegraphics{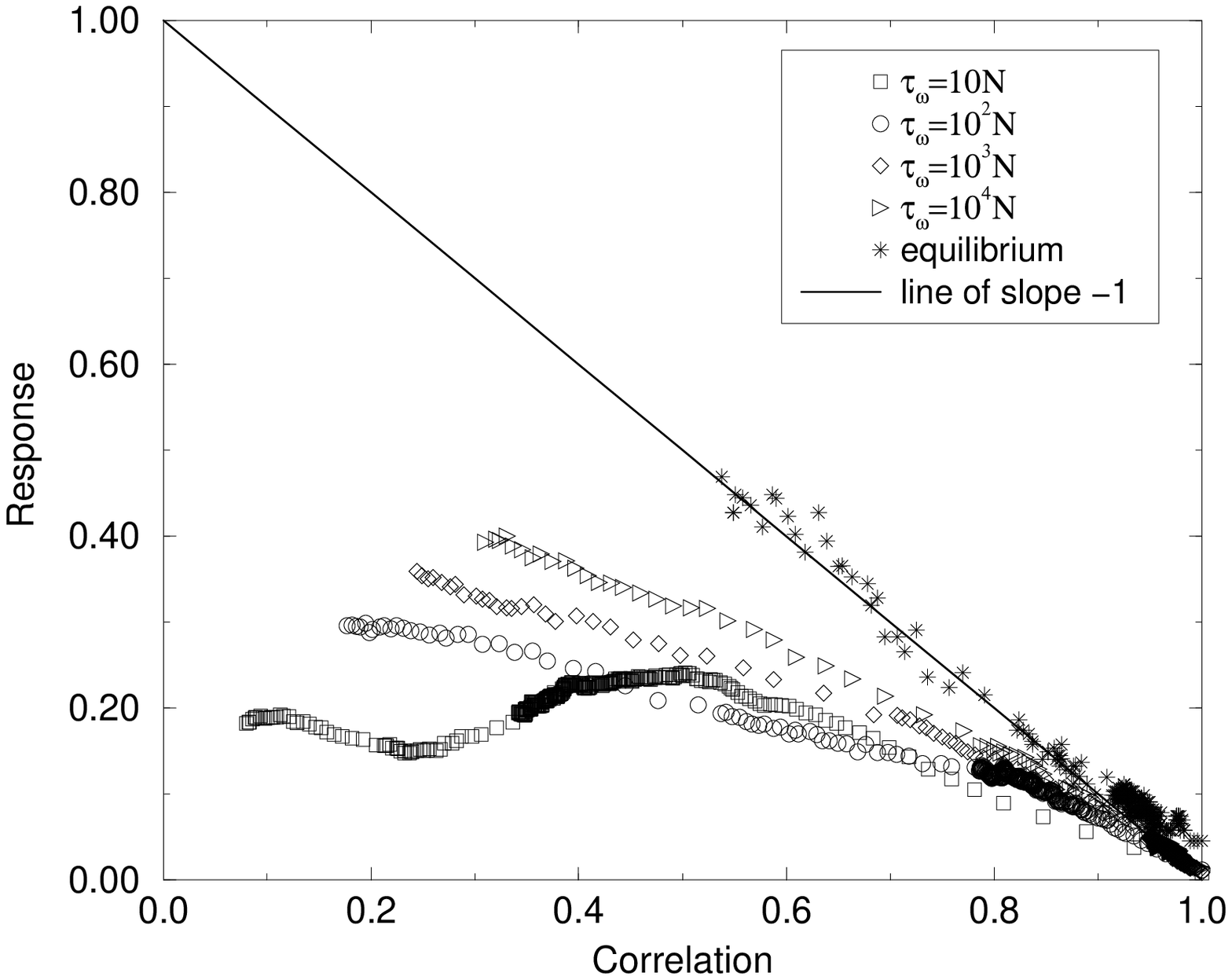}}\label{4.5presp}}
\subfigure[$\beta=6$]{\resizebox{!}{170pt}{\includegraphics{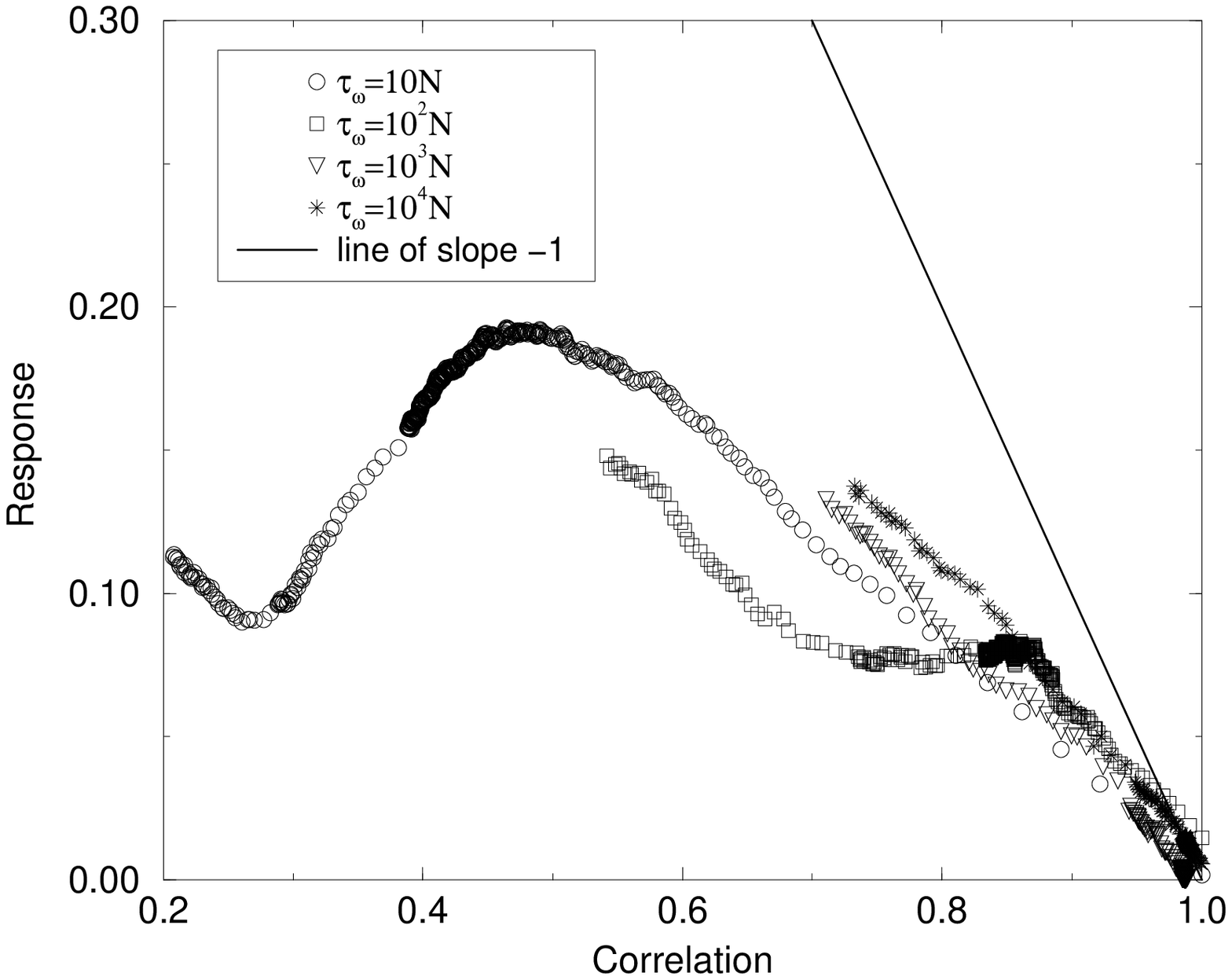}}\label{6presp}}
\caption{\textbf{Parametric plots of the response against the correlation
function for different temperatures and waiting times.}\label{presponse}}
\end{center}
\end{figure}
\begin{figure}[t!]
\subfigure[$\beta=2.5$]{\resizebox{!}{170pt}{\includegraphics{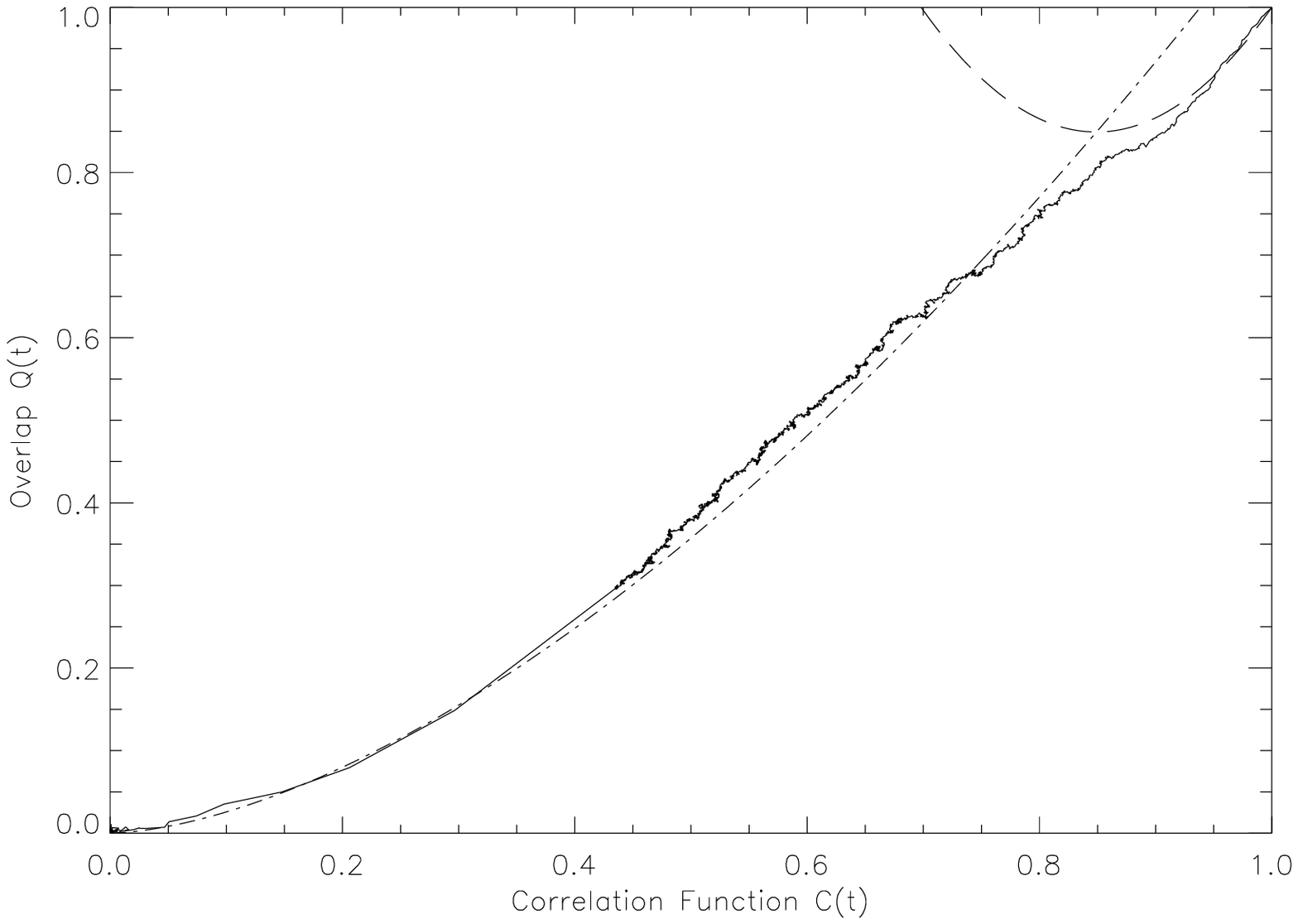}}\label{2.5pq}}
\subfigure[$\beta=3.5$]{\resizebox{!}{170pt}{\includegraphics{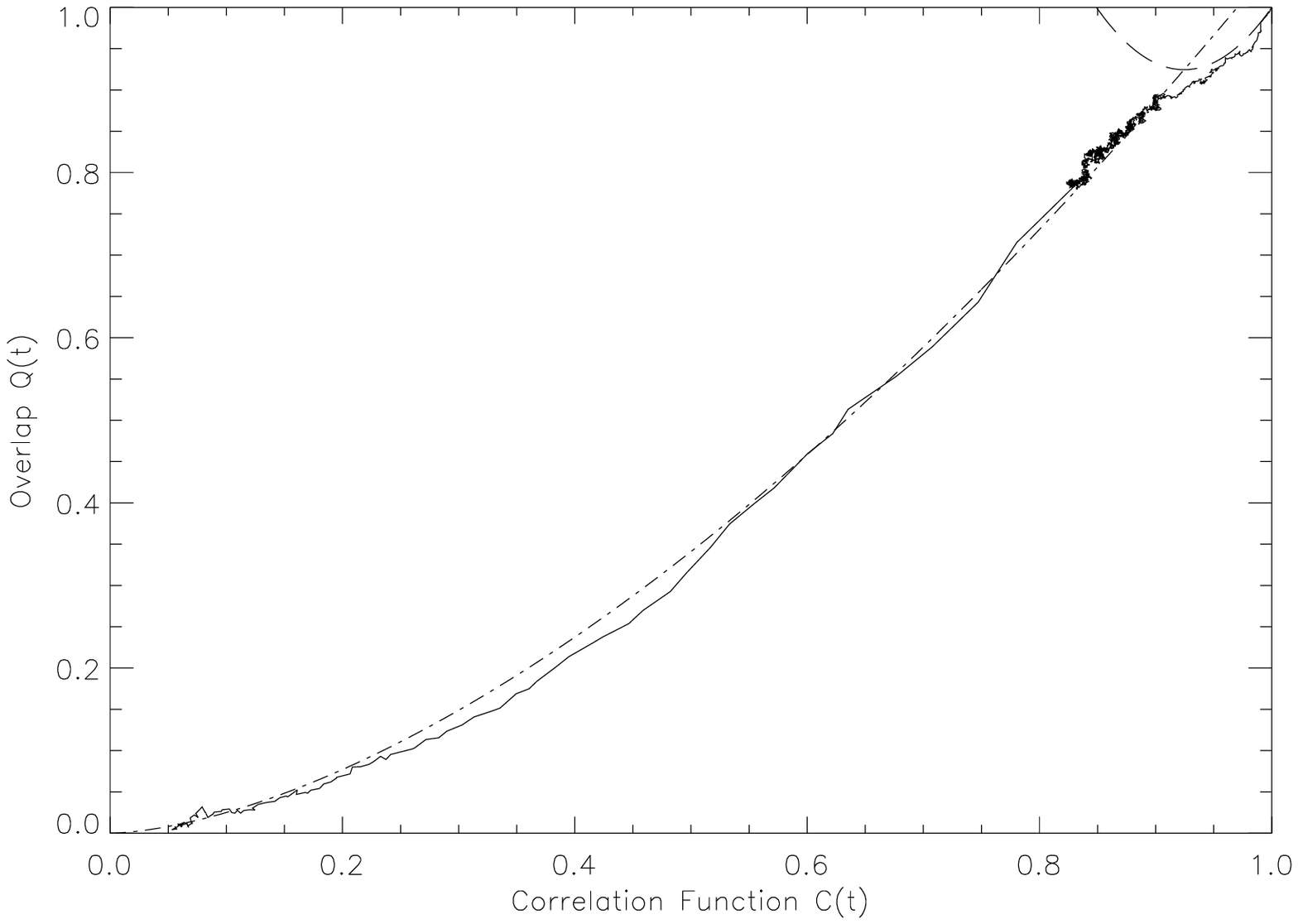}}\label{3.5pq}}
\caption{\textbf{Overlap $\boldsymbol{Q(t)}$ against $\boldsymbol{C(t)}$ in equilibrium.} In each case
the dashed line is the expected short-time behaviour and the dot-dashed
line is the expected long-time behaviour, if equation (\ref{stretch})
holds. The values of $\alpha,\gamma$ are those fitted in the previous section. \label{pqlap}}
\end{figure}

We continue our study of the $D<0$ model by observing the response of the
system to an applied field. As before, a field $h\epsilon_i$ is applied at
time $t_w$, with $\epsilon_i=\pm 1$ randomly at each site. The perturbation this introduces to the energy
is $\Delta E(t) = h \sum_{i=1}^N \epsilon_i \delta_{s(t),0}
\hspace{2pt}\theta(t-t_w)$, where $\theta(t-t_w)$ is again the
Heaviside function. The observable we then measure is the following linear
response function $G(t_w,t_w + t)$:
\begin{equation}
\label{prespeq}
G(t_w,t_w + t) = \frac{1}{h(1-p_{eq}(0))}\left(\frac{\sum_{i=1}^{N}
\epsilon_i \hspace{1pt} \delta_{s(t_w+t),0}}{\sum_{i=1}^{N}\delta_{s(t_w),0}}\right) \cdot
\end{equation}
Again one expects a parametric plot of $-TG(t_w,t_w + t)$ against the
correlation function (defined as in equation (\ref{pcorrel})) to yield
a slope of -1 where the conventional FDT is upheld. 

Figure \ref{presponse} shows such a parametric plot for a variety of
temperatures and waiting times $t_w$. One can see the main features of
the equivalent $D>0$ plots in these figures: FDT is upheld for a time
which increases as $t_w$ increases, and sometime after it is broken
the curves display non-monotonic behaviour. As in the $D<0$ case, this
turnover in the response after long times (i.e. low values of the response
and of $C(t_w,t_w+t)$) is due to the isolated
defects being eliminated from the system, thus decreasing the
response. Note that although this turnover can been seen clearly in
Figure \ref{2.5presp}, one cannot always run the simulations for long
enough to observe
this effect at low temperatures and large waiting times; in fact, it is not observable for any waiting times for
$\beta=6$ (Figure \ref{6presp}). However, in addition to this behaviour, at very low
temperatures and for short $t_w$ we see an intermediate `hump'
appearing before
the turnover due to the isolated defects - see Figures
\ref{4.5presp},\ref{6presp}. This is a consequence of the presence of temperature-dependent dimer
diffusion; blocked dimers exist which take some time to become
mobile and cannot diffuse freely through the system. Instead of being
eliminated before they can make a substantial contribution to the
response, some dimers persist and considerably increase the
response before they are finally removed. This accounts
for the intermediate humps shown for short waiting times in Figures
\ref{4.5presp},\ref{6presp}; at longer waiting times this effect is
imperceptible because the dimers have moved closer to
equilibrium. It is also imperceptible at higher temperatures (lower $\beta$) because the
dimers can move more freely and equilibrate more quickly. 

Let us turn now to the overlap function, defined in this case as:
\begin{equation}
\label{poverlap}
Q_{t_w}(t)=\frac{\sum_{i=1}^N \delta^1_{s_i,0}(t_w + t)
\delta^2_{s_i,0}(t_w + t)}{\sum_{i=1}^N \delta^1_{s_i,0}(t_w)} \cdot
\end{equation}
Recalling that in equilibrium one finds $Q(t)=C(2t)$, we can use the equilibrium overlap to test of our proposed form of
$C(t)$, as given in equation (\ref{stretch}). If this equation holds
we expect to find that for short
times:
\begin{equation}
\label{pQshort}
Q(t) \sim \left(\alpha \mathrm{e}^{-2t/\tau_1} + (1-\alpha) \right) \sim \left(\frac{\left(
C(t) +\alpha -1\right)^2}{\alpha} + (1-\alpha)\right)
\end{equation}
and for long times:
\begin{equation}
\label{pQlong}
Q(t) \sim \left( 1-\alpha \right)
\mathrm{e}^{\left(-2t/\tau_2\right)^\gamma} \sim
\frac{C(t)^{2^{\gamma}}}{\left( 1-\alpha\right)^{2^{\gamma}}} \cdot
\end{equation}
Figure \ref{pqlap} shows the equilibrium results for $\beta=2.5$ and
3.5; the superimposed curves are the expected short and long time
behaviour using the values of $\alpha,\gamma$ obtained from fitting 
the correlation functions with equation (\ref{stretch}). The
theoretical behaviour clearly fits the data very well, lending further
support to equation (\ref{stretch}) as a description of the behaviour
of the equilibrium correlation functions. We also see that $Q(t)$
decays to zero, thus placing the $D<0$ model in the Type II class along
with the $D>0$ model.

\section{Concluding remarks}
We have studied a simple lattice-based spin model which has a non-interacting
Hamiltonian, but constrained dynamics, and find it to exhibit both
glassy behaviour and behaviour typical of diffusion-limited reaction
models. A single parameter $D$ distinguishes two types of ground
state. By choosing $D>0$, one can study a
system with a unique ground state, evolving by way of a number of
annihilation-diffusion processes which are either fast
temperature-independent or slow
temperature-dependent diffusive processes, with the latter slower by a
factor exponential in inverse temperature. We can categorise the fast
processes as: $A + \bar{A} \to \emptyset$, $A + A$ (or $\bar{A} + \bar{A}$)
$\to C + \bar{C}$, $A$ (or $\bar{A}$) $+~C$ (or $\bar{C}$) $\to
\emptyset + C$ (or $\bar{C}$); and the slow processes as: $C + \bar{C}
\to A$ (or $\bar{A}$), where $A$ and $\bar{A}$ are dimers and
anti-dimers, and $C$ and $\bar{C}$ are isolated defects of opposite
sign.  The isolated defects move isotropically, but the dimers and
anti-dimers move anisotropically, and come in three different
`flavours' according to their orientation. The different flavours
can also scatter amongst themselves via the process $A^{\alpha} +
\bar{A^{\alpha}} \to A^{\beta} + \bar{A^{\beta}}$, where
$\alpha,\beta$ label different flavours. Since $A \neq \bar{A}$ and $C
\neq \bar{C}$, $A + \bar{A}$ processes are equivalent to the usual $A
+ B$ processes, and $C + \bar{C}$ are equivalent to $C + D$
processes. In this paper we study the full set of processes
simulationally but only provide a simplified adiabatic theoretical fit. For
$D<0$, the ground state is highly degenerate and the system
evolves according to the fast annihilation-diffusion processes $A + A
\to \emptyset$, $A + A \to C + C$, $A + C \to \emptyset + C$, and the slow
diffusive process $C + C \to A$, where $A$ corresponds to a pair of
zero-spins and $C$ to an isolated zero-spin. Again there are 3
different flavours of $A$ corresponding to the three different
orientations, and these can scatter through $A^{\alpha} + A^{\alpha}
\to A^{\beta} + A^{\beta}$. In this case the movement of $A$'s are
hindered by the background due to the ground state degeneracy, and even
the fast processes have some temperature dependence. As before we have
studied the full dynamics simulationally but only provide a simplified
adiabatic analysis. There is clearly scope for providing a full
analytic theory.

For both the $D>0$ and the $D<0$ case one finds two-step
relaxation, on two different time-scales which are separable and can
be attributed directly to the different processes.  We find that
the energy in the $D>0$ case can be fitted with the sum of two terms,
each behaving like the asymptotic predictions of an $A+B \to
\emptyset$ theory (for the slow processes, $C + \bar{C}
\to A$, but the $A$'s are eliminated on a time-scale which is
negligible compared to that of the slow process, so this behaves like
$C + \bar{C} \to \emptyset$). In the $D<0$ case one cannot
fit the energy adequately in the approach to the intermediate plateau without including induced dimer absorption ($A + C \to
\emptyset ~+~C$) along with the $A + A \to \emptyset$ fast diffusive
processes. We also find that the slow diffusive process does
not behave like a pure $C + C \to \emptyset$ process; this is in part
due to the fact that the dimer diffusion is now temperature
dependent. We have studied the correlation functions for $D$ positive and
negative; in both cases, a naive theory gives a predicted form for
these which fits the data extremely well. Studies of the overlap in
equilibrium serve to reinforce these results. An investigation of the
response function in both cases yields non-monotonic response curves,
and for $D<0$ the temperature-dependent diffusion leads to
more complex results: one finds intermediate humps for very low
temperatures and short waiting times, which can be understood within the
framework of the processes we have already discussed. Non-monotonicity
has also been observed in the response functions of many other models which
involve activated processes \cite{fredrikson,
eisinger,kurchan,  sollich, RitortQ, juanpe,barratloreto, granular}. To
distinguish between Type I (coarsening) and Type II (glassy)
tendencies we have examined an overlap function measuring the temporal
auto-correlation of two independently evolving clones of a
configuration. This demonstrates that the present system is of Type
II, for $D$ both positive and negative.

In this paper we have employed a hexagonal basis for the cell
edges. This is naturally motivated by analogy with a two-dimensional
froth. It also corresponds to the case of the simplest non-trivial
vertices, which have valence three, and consequently is special in that any
cell has two nearest-neighbour cells which are nearest-neighbours of
one another. Extensions are clearly possible, both to higher
valence vertices in two dimensions and to minimal and non-minimal
vertices in higher dimensions, but we do not pursue them here. We
merely note that systems with valence greater than $(d+1)$, where $d$ is
the dimensionality, are more prone to sticking. 

\section{Acknowledgements}
The authors would like to thank J. Cardy, F. Ritort, E. Moro and S. Krishnamurthy for helpful
discussions. LD, DS and JPG would like to thank EPSRC(UK) for financial
support: DS and JPG for research grant GR/MO4426, and LD for research
studentship 98311155. AB acknowledges the support of Marie Curie
Fellowship HPMF-CT-1999-00328. JPG also acknowledges the award of a
Violette and Samuel Glasstone Research Fellowship.

\end{document}